\documentclass[preprint,12pt,authoryear]{elsarticle}

\usepackage[percent]{overpic}

\usepackage{lineno}
\usepackage{amssymb}
\usepackage{amsmath}
\usepackage{amsthm}
\usepackage[colorlinks, allcolors=blue]{hyperref}
\usepackage{placeins}
\usepackage{longtable}
\usepackage{setspace}
\usepackage{tikz}
\usepackage{lscape}
\usepackage{array}
\newcolumntype{M}{>{\centering\arraybackslash}m{2cm}}

\usepackage{subcaption}
\usepackage[margin=25mm]{geometry}
\usepackage[utf8]{inputenc}

\DeclareUnicodeCharacter{2212}{$-$}
\DeclareUnicodeCharacter{03B4}{$\delta$}
\DeclareUnicodeCharacter{2248}{$\approx$}

\graphicspath{{Figures/}}
\journal{Experimental Thermal and Fluid Science}

\begin{document}

\begin{frontmatter}

\title{Investigation of Large Scale Motions in Zero and Adverse Pressure Gradient Turbulent Boundary Layers Using High-Spatial-Resolution PIV}

\author[label1]{Muhammad Shehzad\corref{cor1}}
\author[label1]{Bihai Sun}
\author[label1]{Daniel Jovic}
\author[label2]{Yasar Ostovan}
\author[label2]{Christophe Cuvier}
\author[label2]{Jean-Marc Foucaut}
\author[label3]{Christian Willert}
\author[label1]{Callum Atkinson}
\author[label1]{Julio Soria}

\cortext[cor1]{Corresponding author. Email Address: muhammad.shehzad@monash.edu}

\affiliation[label1]{organization={Laboratory for Turbulence Research in Aerospace \& Combustion (LTRAC), Department of Mechanical and Aerospace Engineering},
            addressline={Monash University}, 
            city={Clayton},
            postcode={3800}, 
            state={Victoria},
            country={Australia}}

\affiliation[label2]{organization={Univ. Lille, CNRS, ONERA, Arts et Metiers Institute of Technology, Centrale Lille, UMR 9014 - LMFL - Laboratoire de M\'ecanique des Fluides de Lille - Kamp\'e de F\'eriet},
            city={Lille},
            postcode={F-59000}, 
            country={France}}
            
\affiliation[label3]{organization={Institute of Propulsion Technology},
            addressline={German Aerospace Center (DLR)}, 
            city={Cologne},
            country={Germany}}

\begin{abstract}
High-spatial-resolution (HSR) two-component, two-dimensional particle-image-velocimetry (2C-2D PIV) measurements of a zero-pressure-gradient (ZPG) turbulent boundary layer (TBL) and an adverse-pressure-gradient (APG)-TBL were taken in the LMFL High Reynolds number Boundary Layer Wind Tunnel. The ZPG-TBL has a momentum-thickness based Reynolds number $Re_{\delta_2} = \delta_2 U_e/\nu = 7,750$ (where $\delta_2$ is the momentum thickness and $U_e$ is the edge velocity), while the APG-TBL has a $Re_{\delta_2} = 16,240$ and a Clauser's pressure gradient parameter $\beta = \delta_1 P_x/\tau_w = 2.27$ (where $\delta_1$ is the displacement thickness, $P_x$ is the pressure gradient in streamwise direction and $\tau_w$ is the wall shear stress). After analysing the single-exposed PIV image data using a multigrid/multipass digital PIV \citep{soria1996investigation} with in-house software, proper orthogonal decomposition (POD) was performed on the data to separate flow-fields into large- and small-scale motions (LSMs and SSMs), with the LSMs further categorized into high- and low-momentum events. The LSMs are energized in the outer-layer and this phenomenon becomes stronger in the presence of an adverse-pressure-gradient. Profiles of the conditionally averaged Reynolds stresses show that the high-momentum events contribute more to the Reynolds stresses than the low-momentum between wall to the end of the log-layer and the opposite is the case in the wake region. The cross-over point of the profiles of the Reynolds stresses from the high- and low-momentum LSMs always has a higher value than the corresponding Reynolds stress from the original ensemble at the same wall-normal location. This difference is up to 80\% in Reynolds streamwise and shear stresses and up to 15\% in the Reynolds wall-normal stresses. Furthermore, the cross-over point in the APG-TBL moves further from the wall than in the ZPG-TBL. By removing the velocity fields with LSMs which contribute significantly to the most energetic POD mode, the estimate of the Reynolds streamwise stress and Reynolds shear stress from the remaining velocity fields is reduced by up to $42 \%$ in the ZPG-TBL. The reduction effect is observed to be even larger (up to 50\%) in the APG-TBL. However, the removal of these LSMs has a minimal effect on the Reynolds wall-normal stress in both the ZPG and the APG cases.

\end{abstract}


\end{frontmatter}

\section{Introduction}
\label{sec:intro}

The identification and characterization of coherent structures in turbulent flows has been an active area of research for decades. In wall-bounded flows, LSMs are defined as coherent patterns that dominate the log-layer and are characterized as alternating regions of high- and low-momentum \citep{kline1967structure,wark1991experimental,ganapathisubramani2005investigation,del2003spectra,hambleton2006simultaneous}. These structures also greatly influence the near-wall region as they superimpose onto the near-wall SSMs and hence, leave their footprints at the wall \citep{hutchins2007evidence}. Furthermore, LSMs in the log-layer cause an amplitude modulation of the SSMs in the near-wall region \citep{mathis2009large,harun2013pressure,hutchins2007large}. These regions of LSMs are elongated up to the order of $20\delta$ in the streamwise direction \citep{hutchins2007evidence}, where $\delta$ is the boundary layer thickness. They are distinguished from the very thin low-speed streaks in the buffer layer at high Reynolds numbers due to their much larger wall-normal extent \citep{liu2001large}. A slightly inclined and elongated streamwise velocity correlation function implies that the LSMs are significant contributors to the streamwise turbulent kinetic energy (TKE) as observed by both \citet{grant1958large} and \citet{townsend1958turbulent}.

While there have been a number of studies dealing with the statistical nature of TBLs under the influence of a pressure gradient (see \cite{nagano1993effects,spalart1993experimental,skaare1994turbulent,krogstad1995influence,fernholz1998effects,na1998direct,bourassa2009experimental,kitsios2017direct,cuvier2017extensive,sekimoto2019outer,senthil2020analysis} among others), literature reveals that only a few characterize the coherent structures \citep{lian1990visual,zhou1997effect,houra2000effects,lee2009structures,drozdz2011detection,kitsios2017direct,sekimoto2017intense,senthil2020analysis} and fewer characterize the LSMs \citep{harun2013pressure,hain2016coherent,bross2019interaction}. Using spectral analysis, \citet{harun2013pressure} studied the effect of pressure gradients on the LSMs in a ZPG-TBL at a $Re_{\delta_2} = 8,160$ and in an APG-TBL at a $Re_{\delta_2} = 12,030$ and a $\beta = 1.74$. They observed that the large scales are energized in the entire APG-TBL and that the amplitude modulation of near-wall small scales by the LSMs increases with an increasing pressure gradient. 

\citet{bross2019interaction} studied the interaction of coherent structures in the near-wall region of an APG-TBL using high-resolution time-resolved 2D and 3D particle tracking velocimetry (PTV). They reported that the high-momentum LSMs in the buffer- and log-layer region were associated with positive fluctuating wall-shear stress $+\tau'_w$ and that low-momentum LSMs were associated with $-\tau'_w$. Thus, the LSMs in the log-layer significantly manipulate the wall-shear stress, $\tau'_w$. They also presented a three-layered model of uniform-momentum-zones (UMZs) in a TBL, of which the first zone is associated with the viscous sublayer, the second is formed by the high- and low-speed streaks in the buffer-layer and the third is generated by the high- and low-momentum LSMs in the log-layer. When these low- and high-momentum LSMs coincide with ejections or sweeps (i.e. Q2 and Q4 events according to the quadrant analysis of \citet{wallace1972wall}), the number of UMZs strongly increases or decreases, respectively \citep{paakkari2018uniform}. Hence, \citet{bross2019interaction} conclude that the momentum of log-layer LSMs and their footprint on the wall are related to the number of UMZs in the instantaneous velocity fields. 

Proper Orthogonal Decomposition (POD) has been used as a tool to study the coherent structures, including LSMs since its first application in fluid dynamics by \citet{lumley1967structure}. It is a generalization of the conventionally used Fourier power spectral analysis and is used to investigate the TKE distribution as a function of scale in a TBL flow that is inhomogeneous in the streamwise direction \citep{liu2001large}. POD has been used to study the randomly distributed counter-rotating eddies as LSMs in a turbulent pipe flow by \citet{bakewell1967viscous} at a $Re_D = U_bD/\nu = 8,700$, where $U_b$ is the bulk velocity and $D$ is the pipe diameter. \citet{liu2001large} used POD to evaluate the scales contributing towards the events that produce TKE and Reynolds shear stress from the 2D data of two different channel flows at a $Re_h = U_bh/\nu = 5,378$ and $29,935$ (where $h$ represents the channel half-height) and concluded that the LSMs contain a large fraction of the Reynolds streamwise stress component and a small fraction of Reynolds wall-normal stress component. \citet{wu2014study} studied the LSMs in a TBL with a zero pressure gradient (ZPG) at two different configurations, $Re_{\delta_2} = 8200$ and $12000$, and performed POD to establish a connection between the first two dominant POD modes and the instantaneous large scale structures. The authors concluded that the Reynolds streamwise stress, Reynolds shear stress and spatial velocity correlation functions are reduced without the LSMs that significantly contribute to the dominant first POD mode.

By performing POD on a dataset, the LSMs are characterized as structures whose contribution to the dominant spatial mode is above a threshold. The classification of LSMs into high- and low-momentum events is based on the nature of the first POD mode. Conditional averaging of velocity fields based on this classification leads to the analysis of the contribution of the extreme events to the turbulent statistics. \citet{guemesexperimental} studied the LSMs in a ZPG-TBL by using POD to investigate the effect of high- and low-momentum events on turbulent statistics. They observed that the high-momentum events have a larger influence on the mean flow and Reynolds stresses near the wall when compared to the low-momentum events, and this effect decreases with an increasing distance from the wall. Conversely, the low-momentum events had a weaker contribution to the inner peak and a stronger contribution to the outer peak in Reynolds streamwise stress profile. To the best of authors' knowledge, the effect of the high and low-momentum events on the turbulent statistics in TBL under the influence of an adverse pressure gradient has not yet been investigated. 

Earlier studies on the LSMs in wall-bounded flows were carried out using either single-point measurements or low-spatial-resolution 2C-2D PIV measurements. To investigate the effect of high- and low-momentum events near the wall, HSR measurements are of immense importance. In the present paper, HSR 2C-2D PIV measurements have been used to investigate the LSMs and the effect of high- and low-momentum events on the turbulent statistics in both the ZPG- and the APG-TBLs. This paper is organized in the following manner: The methodology of the POD on the instantaneous fluctuating flow field and classification of LSMs into high- and low-momentum events is presented in Section \ref{sec:POD_and_LSM}. Section \ref{sec:experiemnt} describes the experimental details of the HSR 2C-2D PIV measurements. The first- and second-order statistics, conditionally averaged statistics and sensitivity analysis of different threshold limits along with discussion are presented in Section \ref{sec:results}. Lastly, the conclusions are presented in Section \ref{sec:conclusion}. 

Throughout this paper, we take $x, y$ and $z$ as the streamwise, wall-normal and spanwise directions, respectively. The instantaneous, mean and fluctuating velocities in the $x$ directions are referred to as $u$, $U$ and $u'$, respectively. Accordingly, the velocities in $y$ and $z$ are represented by $v$ and $w$.

\section{Identification of the Large Scale Motions using POD}
\label{sec:POD_and_LSM}
\subsection{Proper Orthogonal Decomposition (POD)}
\label{sec:POD}

POD using a snapshot method was introduced by \citet{sirovich1987turbulence}, briefly discussed in \citet{taira2017modal} and used in this study to extract modes based on optimizing the mean square of the fluctuating velocity. The method is briefly presented below.

Consider a set of velocity fields $\overline{\overline{\boldsymbol{u'}}}(x,t)$, 

\begin{equation} 
\overline{\overline{\boldsymbol{u'}}}(x,t) = [\boldsymbol{u'}(x,t_1) \;\;\; \boldsymbol{u'}(x,t_2) \;\;\; ... \;\;\; \boldsymbol{u'}(x,t_N)]  \in \mathbb{R}^{M\times N}, \;\;\; M \gg N
\end{equation}
where $N$ is the number of snapshots (i.e. velocity fields) and M is twice the number of grid points in each snapshot and

\begin{equation}
\boldsymbol{u'}(x,t) = 
\begin{bmatrix}
u'(x,t) \\
v'(x,t) \\
\end{bmatrix} 
\end{equation}

$\overline{\overline{\boldsymbol{u'}}}(x,t)$ can be decomposed by POD in the following manner 

\begin{equation}
    \overline{\overline{\boldsymbol{u'}}}(x,t) = \sum_{i=1}^{N} a_i(t) \phi_i(x)
\end{equation}
where $\phi_i(x)$ is the $i$th spatial mode and $a_i(t)$ is the set of the corresponding temporal coefficients.

Let $\boldsymbol{X} = \overline{\overline{\boldsymbol{u'}}} $. In the regular POD, 

\begin{equation}
    \boldsymbol R \phi_i = \lambda_i \phi_i, \; \;  \phi_i \in \mathbb{R}^{M}, \; i=1,2,\; ... \;,N
\end{equation}
where $\lambda_i$ represents the eigenvalue of the $i$th mode and $\boldsymbol R$ is the covariance matrix of vector $\boldsymbol u'(x,t)$ such that

\begin{equation}
    \boldsymbol R = \boldsymbol X \boldsymbol X^T, \; \; \boldsymbol R \in \mathbb{R}^{M \times N}
\end{equation}

In snapshot-POD, the matrix $\boldsymbol{X}^T \boldsymbol{X}$ is used which is much smaller in size as compared to the regular POD matrix $\boldsymbol{X} \boldsymbol{X}^T$ and yet has the same nonzero eigenvalues \citep{sirovich1987turbulence}. Hence we can write

\begin{equation}
    \boldsymbol{X}^T \boldsymbol{X} \psi_i = \lambda_i \psi_i, \;\; \psi_i\in \mathbb{R}^N,
\end{equation}

With the eigenvector (i.e. vector of temporal coefficients) $\psi_i$ and eigenvalue $\lambda_i$, the corresponding spatial mode $\phi_i $ can be computed as 

\begin{equation}
    \phi_i = \boldsymbol{X} \psi_i \frac{1}{ \sqrt{\lambda_i}} 
\end{equation}

This can also be written as 
\begin{equation}
    \boldsymbol{\Phi} = \boldsymbol{X} \boldsymbol{\Psi} \boldsymbol{\Lambda}^{-1/2} 
\label{eq:SPOD_eq}
\end{equation}

\noindent where the columns of $\boldsymbol{\Phi}$ are the vectors of the spatial modes ($\boldsymbol \Phi = \big[\phi_1 \; \phi_2 \; ... \; \phi_N \big] \in \mathbb{R}^{M \times N}$), the columns of $\boldsymbol{\Psi}$ are the vectors of temporal coefficients corresponding to each POD mode ($\boldsymbol \Psi = \big[\psi_1 \; \psi_2 \; ... \; \psi_N \big] \in \mathbb{R}^{N \times N}$) and $\boldsymbol{\Lambda}$ is the vector of  eigenvalues corresponding to each POD mode ($\boldsymbol \Lambda = \big[\lambda_1 \; \lambda_2 \; ... \; \lambda_N \big] \in \mathbb{R}^{N}$).

The TKE equals to half of the sum of the eigenvalues, i.e.
\begin{equation}
    k = \frac{1}{2} \overline{\boldsymbol {u'}^2} = \frac{1}{2} \sum_{i=1}^{N} \lambda_i
\end{equation}

The singular value decomposition (SVD) is defined as
\begin{equation}
    \overline{\overline{\boldsymbol{u'}}} = \boldsymbol{\Phi \Sigma \Psi^T}
    \label{eq:SVD}
\end{equation}

Equation \ref{eq:SPOD_eq} can be written in the form of equation \ref{eq:SVD} such that $\boldsymbol{\Sigma} = \boldsymbol{\Lambda}^{1/2}$ and $\boldsymbol \Psi^{-1} = \boldsymbol \Psi^T$ (as $\boldsymbol \Psi$ is an orthogonal matrix), and hence the current POD problem can be solved as an SVD problem \citep{taira2017modal}.

\subsection{Classification of the Large Scale Motions}
\label{sec:classification}

As described in \citet{wu2014study}, the velocity fields dominated by LSMs are identified as the fields for which the magnitude of the temporal coefficients $\psi_{i_j}$ are beyond a threshold $K \sigma_{\psi_i}$, where $\sigma_{\psi_i}$ is the standard deviation of the temporal coefficients corresponding to the $i$th POD mode, and $K = \{1.0,1.5,2.0\}$ is a threshold factor. As the first POD mode contains the largest contribution of the TKE, we take $i=1$ to separate the flow field into the velocity fields with LSMs and SSMs. 

\begin{equation}
\begin{aligned}
F(|\psi_{i}| > K \sigma_{\psi_i}) \longrightarrow F(LSM), \\
F(|\psi_{i}| < K \sigma_{\psi_i}) \longrightarrow F(SSM)
\label{eq:classification_0}
\end{aligned}
\end{equation}

The LSMs are further classified into the high-momentum ($u'>0$) and low-momentum ($u'<0$) events based on the nature of the first POD mode \citep{wu2014study,guemesexperimental}. If the streamwise component of the first mode is positive, i.e. $u'_{\phi_1}>0$ , the snapshots of the flow field $F$ with their temporal coefficients $\psi_1$ larger than $K \sigma_{\psi_1}$ are identified as those with dominant high-momentum events ($H^+$) and those with $\psi_1$ smaller than $-K \sigma_{\psi_1}$ are identified as those with dominant low-momentum events ($L^+$). 

\begin{equation}
\begin{aligned}
F(\psi_1 > K \sigma_{\psi_1}) \longrightarrow H^+, \\
F(\psi_1 < - K \sigma_{\psi_1}) \longrightarrow L^+
\label{eq:classification_1}
\end{aligned}
\end{equation}

Conversely, if $u'_{\phi_1}<0$, the snapshots of the flow field $F$ with $\psi_1 > K \sigma_{\psi_1}$ are identified as fields with dominant low-momentum events ($L^-$), and those with $\psi_1< -K \sigma_{\psi_1}$ as fields with dominant high-momentum events ($H^-$)
\begin{equation}
\begin{aligned}
F(\psi_1 > K \sigma_{\psi_1}) \longrightarrow L^-, \\
F(\psi_1 < -K \sigma_{\psi_1}) \longrightarrow H^-
\label{eq:classification_2}
\end{aligned}
\end{equation}

The $+$ and $-$ superscripts have been adopted to differentiate between the nature of the first POD mode and subsequent classification in the two data-sets (i.e. of ZPG- and APG-TBLs). The basis of the selection of the optimal value of $K$ is described in section \ref{sec:POD_modes} and compared with the other values to analyse their effect on the conditionally averaged turbulent statistics in section \ref{sec:sensitivity_analysis}.

\section{Experimental methodology}
\label{sec:experiemnt}
\subsection{Facility and Apparatus}
\label{subsec:facility}

\begin{figure}
\begin{overpic}[width=\textwidth]{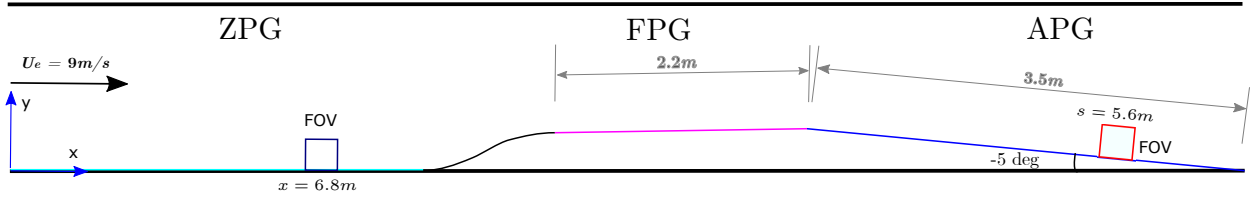}
\put(22,-1){\tiny$x=6.8m$}
\put(86,5){\rotatebox{355}{\tiny$s=5.6m$}}
\end{overpic}
\caption{Schematic of the test section in the LML Wind Tunnel. Figure adapted from \cite{cuvier2017extensive}.}
\label{fig:LML_wind_tunnel}
\end{figure}

High-spatial-resolution 2C-2D PIV images were taken in the $x-y$ plane in the LMFL High-Reynolds-Number Boundary Layer Wind Tunnel at Laboratoire de Mécanique des Fluides de Lille (LMFL), Lille, France. This facility has a $2m$-wide, $1m$-high and $20.6m$-long test section. A schematic diagram of the wind tunnel facility is shown in figure \ref{fig:LML_wind_tunnel} with the three different sections indicated; with a zero-pressure-gradient, with a favourable-pressure-gradient (FPG) and finally with an adverse-pressure-gradient. Previously in the EuHIT experiment \citep{cuvier2017extensive}, 2C-2D PIV measurements of an FPG-TBL with four sCMOS cameras at four stations and an APG-TBL with 16 sCMOS cameras in a $3.466m$ long continuous field of view (FOV) were obtained to characterize the high Reynolds number FPG- and APG-TBL developing over considerably long regions of the test section before entering the FPG region. The details of the EuHIT experiment are reported in \citet{cuvier2017extensive}. 2C-2D Planar PIV and 3C-3D Stereo-PIV (SPIV) measurements of the ZPG-TBL were taken at $x=3.2m$ and $x=6.8m$ respectively to characterize the inlet conditions of the TBL. 

To study the dynamics of the TBL in the near-wall and log regions, HSR 2C-2D PIV measurements of a ZPG-TBL were taken at $x=6.8m$, whereas the measurements of the APG-TBL were taken at $s=5.6m$, where $s$ is the curvilinear coordinate along the ramp surface with $s=0$ at the beginning of the ramp. The inflow velocity for both measurements is $9m/s$. A 29 Megapixel Imperx Bobcat B6640 camera was used to record these images. The camera has a CCD sensor with 6576 x 4384 pixels and a pixel size of $5.5 \mu m$. The image magnification was $12.684\mu m/px$ for ZPG-TBL and $12.726\mu m/px$ for APG-TBL. The FOV was $ (56.190 \times 46.449) mm$ for the ZPG-TBL and $(47.213 \times 49.656) mm$ for the APG-TBL. Due to the limited number of high-resolution cameras available, HSR 2C-2D PIV measurements of the whole APG region in the LMFL wind tunnel were not possible. Therefore, APG-TBL measurements were taken only on at one station in the APG region. The FOV was illuminated by a dual cavity, frequency-doubled Innolas Nd:YAG laser for ZPG-TBL and BMI laser for APG-TBL with the maximum energy of $150mJ$ and $200mJ$ per pulse at a wavelength of $532nm$. The laser sheet was created by a combination of converging and diverging lenses to reduce the thickness in the out-of-plane direction and expand the laser beam to form a sheet in the in-plane direction. Two cylindrical lenses of focal lengths of $f=-1000mm$ at the output of the laser and $f=+400mm$ at about $400mm$ upstream the FOV were used to create the laser sheet for the ZPG-TBL measurements. For the APG-TBL measurements, a cylindrical lens of $f=-40mm$ and a spherical lens of $f=1500mm$ were used to create the required laser sheet. The thickness of the laser sheet across the field of view was measured to be about $400\mu m$ for the ZPG-TBL and about $200\mu m$ for the APG-TBL. The laser sheet for the ZPG-TBL was introduced from underneath the glass-floor of the wind tunnel at $x=6.8m$. For the APG-TBL, the laser sheet was introduced from the bottom at a location downstream of the APG region but mirrored to align its bottom edge coincident with the $-5^{\circ}$ inclined plate. The elapsed time between the two laser pulses $\Delta t$ was set to be $30 \mu s$ for both TBLs to achieve an optical maximum particle image displacement of $20px$ on the sensor. The flow was seeded with droplets of a water-glycol mixture with a mean diameter of around $1\mu m$, using a smoke generator. For each TBL, over $10,000$ image pairs were recorded. As the purpose was to take HSR 2C-2D PIV images, the measurements for both TBLs could not capture the whole boundary layer, but instead only a region from the wall up to the beginning of the wake region. The boundary layer parameters for both TBLs which could not be computed from the current HSR 2C-2D PIV data, have been adopted from \citet{cuvier2017extensive} as presented in table \ref{tab:lml_bl_parameters}.

\begin{table}
\begin{center}
\caption{Boundary layer parameters from EuHIT experiment. Source: \cite{cuvier2017extensive}}
\label{tab:lml_bl_parameters} 
\begin{tabular}{p{1cm}p{1.2cm}p{1.2cm}p{1.2cm}p{1.2cm}p{1.1cm}p{1.1cm}p{0.7cm}}
\hline \hline\noalign{\medskip}
TBL	&$U_e(m/s)$	&$\delta(mm)$ 	&$\delta_1 (mm)$ 	&$\delta_2(mm)$  	&$H$		&$Re_{\delta_2}$ &$\beta$\\
\noalign{\smallskip}\hline\noalign{\medskip}
ZPG	& 9.64			&102	 &16.4	 	&12.0	  	&1.37		&7,750 		&-   \\
APG	& 11.59		&175	 	&33.5	 	&21.0	  	&1.45		&16,240 		&2.27 \\
\hline
\end{tabular}
\end{center}
\end{table}

\subsection{Distortion Correction}
\label{subsec:distortion_correction}

When measuring with large imaging sensors, lens distortions are introduced into the PIV images. These are usually radial distortions which are a minimum at the centre of the image and a maximum in the corners. To correct for these distortions, the PIV images are dewarped using a second-order rational function (R22) \citep{willert1997stereoscopic} given below:

\begin{equation*}
\begin{aligned}
\hat{x}_{i_j} = \frac{a_{11} \hat{x}_{o_j} + a_{12} \hat{y}_{o_j} + a_{13} + a_{14} \hat{x}_{o_j}^2 + a_{15} \hat{x}_{o_j} \hat{y}_{o_j} + a_{16} \hat{y}_{o_j}^2}{a_{31} \hat{x}_{o_j} + a_{32} \hat{y}_{o_j} + a_{33} + a_{34} \hat{x}_{o_j}^2 +  a_{35} \hat{x}_{o_j} \hat{y}_{o_j} + a_{36} \hat{y}_{o_j}^2} ,\\
\end{aligned}
\end{equation*}

\begin{equation}
\begin{aligned}
\hat{y}_{i_j} = \frac{a_{21} \hat{x}_{o_j} + a_{22} \hat{y}_{o_j} + a_{23} + a_{24} \hat{y}_{o_j}^2 + a_{25} \hat{x}_{o_j} \hat{y}_{o_j} + a_{26} \hat{y}_{o_j}^2 }{a_{31} \hat{x}_{o_j} + a_{32} \hat{y}_{o_j} + a_{33} + a_{34} \hat{x}_{o_j}^2 + a_{35} \hat{x}_{o_j} \hat{y}_{o_j} + a_{36} \hat{y}_{o_j}^2},\\
\end{aligned}
\label{eq:dewarping_function}
\end{equation}
$$a_{33} = 1$$
    
\noindent where $(\hat{x}_{i_j},\hat{y}_{i_j})$ and $(\hat{x}_{o_j},\hat{y}_{o_j})$ are coordinates of the image points and the object points in the centred and normalized image space $\{\hat P_{I}\}$ and object space $\{\hat P_{O}\}$ respectively. $j = [ 1,2,3, \; ... \;,N_m]$ represents a pair of the corresponding image and object points for an individual marker and $N_m$ is the total number of markers in the calibration target image. 

The locations of all the markers are obtained with a sub-pixel accuracy by cross-correlating the calibration image with a custom marker template and using a peak finding algorithm. This creates the image space, $\overline{P_{I}}$. The approximate magnification $M$ is calculated from the average distance between the four markers in the centre of the image space because distortion is minimum there. $M$ is then used to create the object space $\{\overline{P_{O}}\}$ with its origin coincident with the origin of the image space. The centred and normalized image space and object space are obtained from their original counter parts $\overline{P_{I}}$ and $\overline{P_{O}}$ as follows:

\begin{equation}
\{\hat{P}_{I}\}  = \frac{\{\overline{P_{I}}\} - \overline{P_{i_c}}}{x_{o_{max}}}
\end{equation}
\begin{equation}
\{\hat{P}_{O}\}  = \frac{\{\overline{P_{O}}\} - \overline{P_{o_c}}}{x_{o_{max}}}
\end{equation}

\noindent where $\overline{P_{i_c}}$ and $\overline{P_{o_c}}$ are the image and object points at the middle of the image space and object space respectively and $x_{o_{max}}$ is the largest $x$ coordinate in the object space. 

\begin{figure}
  \includegraphics[trim={0.0cm 0cm 0cm 0cm},clip,width=0.9\textwidth]{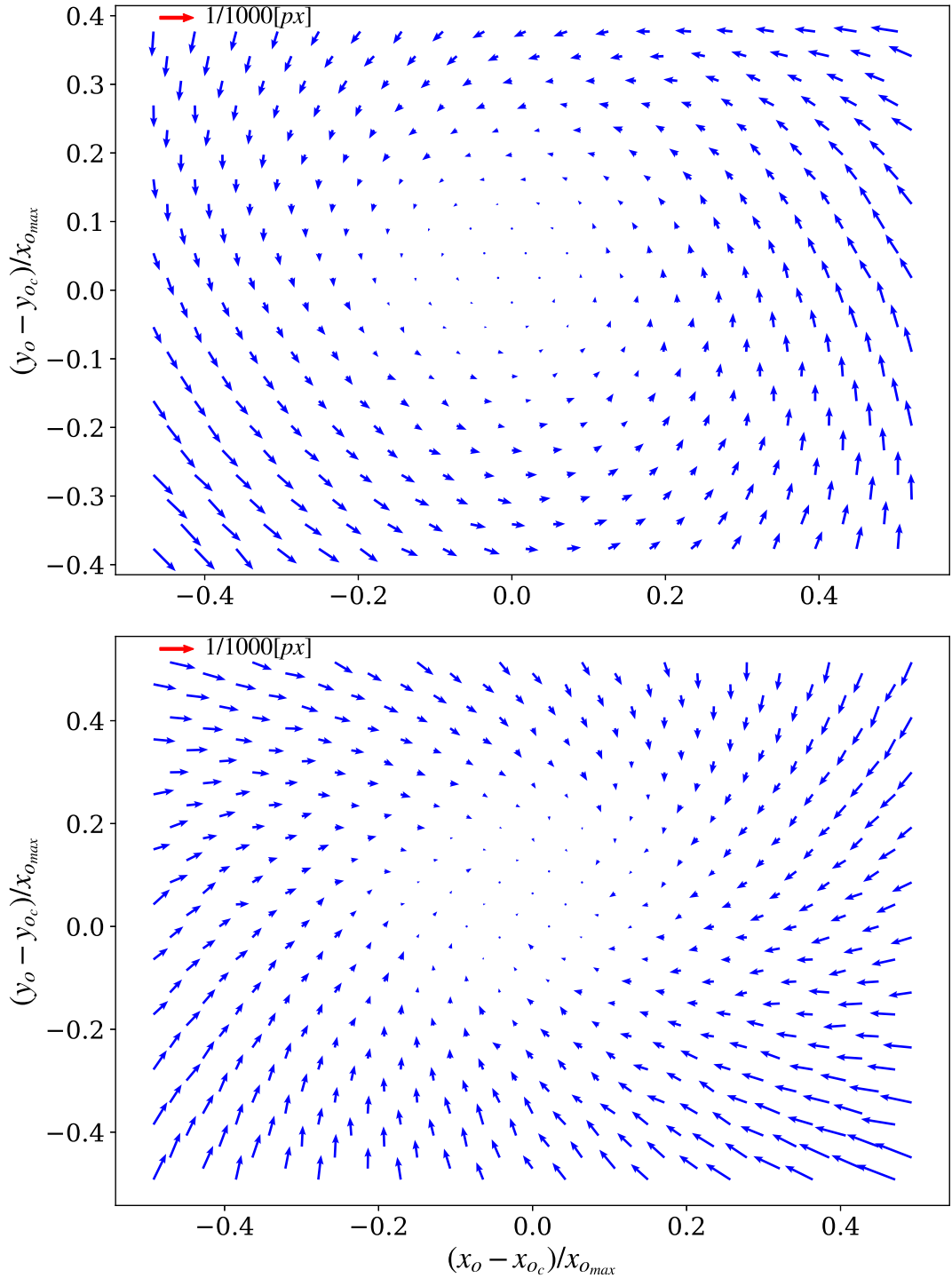}
\caption{Difference of the centred and normalized object  and image points with blue arrows pointing from the object points towards the corresponding image points for the ZPG on top and APG on bottom. The number of arrows ({\em i.e.} markers) have been down-sampled to help visualize it better.}
\label{fig:imgObjPoints_centred_normalized_bothTBL}       
\end{figure}

The difference of the centred and normalized object and image points is highlighted in Fig. \ref{fig:imgObjPoints_centred_normalized_bothTBL} where the arrows extend from the object points towards their corresponding image points. The maximum difference of the image points and object points are observed to be $6.73px$ and $5.68px$ in the ZPG- and APG-TBLs PIV images respectively. As shown in the figure, the distortion in ZPG-TBL images is a combination of a solid body rotation and pincushion type while in APG-TBL, it is mostly the pincushion type distortion. 

Complete details of the distortion correction method are given in \cite{bihai2021distortion}. The characteristic parameters required to compute the dewarping coefficients $$[a_{11}, a_{12}, \; ... \;, a_{16}, a_{21}, a_{22}, \; ... \;, a_{26}, a_{31}, a_{32}, \; ... \;, a_{36}]$$ of equation \ref{eq:dewarping_function} for distortion correction of the current HSR ZPG- and APG-TBLs measurements are listed in table \ref{tab:distorion_parameters}.

\begin{table}[tbph]
\begin{center}
\caption{The characteristic parameters to compute the dewarping coefficients of the equation \ref{eq:dewarping_function}.}
\begin{tabular}{ccccc}
\hline \hline\noalign{\medskip}
Property  & Units  & ZPG & APG \\
\noalign{\smallskip}\hline\noalign{\medskip}
$\overline{\Delta x}_o, \overline{\Delta y}_o $  & ($mm$) & $1\pm0.001$ & $1\pm0.001$ \\ 
$M$ 	&($px/mm$) &78.839 &78.580 \\ 
$\overline{P}_{I_c}$ & ($px$) & (2108.7, 1697.7)  &(1865.1, 1860.8)  \\ 
$\overline{P}_{O_c}$ & ($px$)   &(2105.8, 1700.6)  &(1863.7, 1858.0)	 \\
$x_{o_{max}}$ & ($px$)  &4392.1	  &3671.3	\\
$\hat{P}_{I_{min}}$ & ($px$)  &(-0.467, -0.378)  &(-0.493, -0.493)		\\
$\hat{P}_{I_{max}}$  & ($px$)  &(0.521,  0.395)  &(0.493 ,  0.514) 	\\
 $\hat{P}_{O_{min}}$  & ($px$)  &(-0.467, -0.377)  &(-0.492, -0.492)   \\
 $\hat{P}_{O_{max}}$  & ($px$)  &(0.521,  0.395)  &(0.492,  0.514)\\
\hline
\label{tab:distorion_parameters} 
\end{tabular}
\end{center}
\end{table}

\subsection{PIV Analysis}
\label{subsec:piv_analysis}

The PIV images were analysed using a multigrid/multipass digital PIV \citep{soria1996investigation} with in-house software. The parameters of the PIV analysis are given in terms of viscous units in table \ref{tab:piv_parameters_viscous}. In order to compare the first and second-order statistics in the outer layer, profiles of the ZPG- and APG-TBLs from the EuHIT experiment measured at the same streamwise stations and their PIV analysis parameters, are presented in section \ref{sec:flow_stats}. These have been adopted from \cite{cuvier2017extensive}. The current measurements are approximately $3$ times more resolved in the streamwise direction for both of the TBLs while being about $27$ and $8$ times more resolved in wall-normal direction for the ZPG- and the APG-TBLs respectively when compared to their EuHIT counterparts from \citet{cuvier2017extensive}. As the current PIV measurements have significantly higher resolution in the wall-normal direction, the wall shear stress $\tau_w = \mu \left(\frac{dU}{dy}\right)_{y=0}$ and friction velocity $u_\tau = \sqrt{\tau_w/\rho }$ for both the TBLs are measured directly from wall-normal gradient of the mean streamwise velocity in the viscous sublayer at every streamwise grid-point. The mean values of the measured friction velocities are given in table \ref{tab:bl_parameters_current}. These values are in good agreement with those measured in \cite{cuvier2017extensive} with a deviation of about $1\%$.

\begin{table}[h]
\begin{center}
\caption{PIV analysis parameters. Source for EuHIT data: \cite{cuvier2017extensive}}
\begin{tabular}{ccccc}

\hline \hline\noalign{\medskip}
Measurement	&  ZPG & APG   &   ZPG(EuHIT)  & APG(EuHIT) \\
\hline
Inflow Velocity ($m/s$)  & 9 & 9 & 9 & 9 \\ 
Viscous length scale $l^+$ ($\mu m$)	 &42   &44   &42   &44  \\	
FOV ($l^+ \times l^+$)   &$1,492 \times 1,090$     &$1,456 \times 1,115$  &$7,085 \times 4,265$	 &$78,847 \times 5801$  \\
Grid spacing ($l^+ \times l^+$)  &$5 \times 1$  &$5 \times 2$	   &$26 \times 39$		 &$34 \times 34$ \\

IW  size ($l^+ \times l^+$) &$19 \times 2$	   &$19 \times 7$		&$57 \times 57$  	&$58 \times 58$	\\
Number of samples   &10,479   &10,479   &10,000	  &30,000	\\
Vector field size 	&$269 \times 593$    &$224 \times 633$ 	 	&$299 \times 180$  	&$3,250 \times 238$  \\

\hline
\label{tab:piv_parameters_viscous} 
\end{tabular}
\vspace{1em}
\end{center}
\end{table}

\begin{table}[h]
\begin{center}
\caption{Boundary layer parameters of the current HSR 2C-2D PIV measurements.}
\begin{tabular}{cc}
\hline \hline\noalign{\medskip}
Flow condition	&$u_\tau(m/s)$ \\
\noalign{\smallskip}\hline\noalign{\medskip}
ZPG		 &0.35 \\
APG	     &0.33\\
\hline
\label{tab:bl_parameters_current} 
\end{tabular}
\end{center}
\end{table}

\section{Results and discussion}
\label{sec:results}

\subsection{First- and second-order statistics}
\label{sec:flow_stats}

\begin{figure}[h]
\begin{center}
  \includegraphics[width=0.95\textwidth]{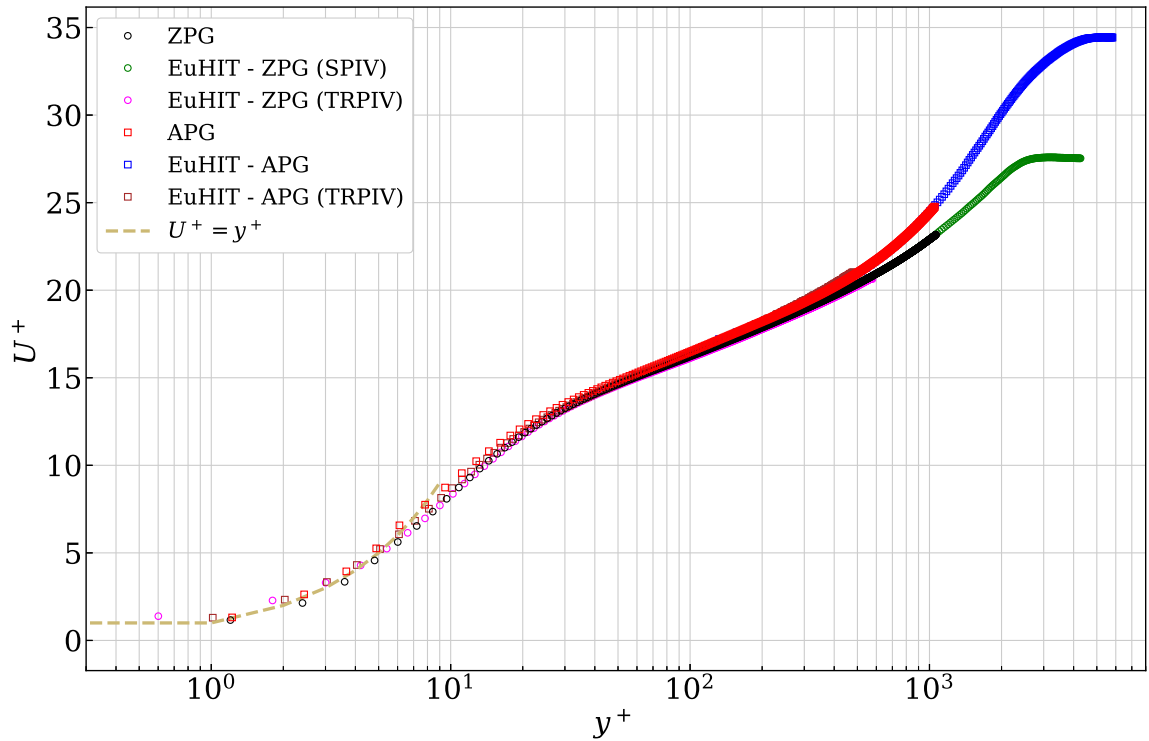}
\caption{Mean velocity profiles.}
\label{fig:MVP}
\end{center}
\end{figure}

\noindent Mean streamwise velocity profiles of ZPG- and APG-TBLs normalized by the viscous units are shown in Fig. \ref{fig:MVP}. The ZPG-TBL has a longer log-law region than that of the APG-TBL. For comparison, the outer-layer mean velocity profiles of the ZPG- and APG-TBLs from the EuHIT experiment have also been included which are consistent with their respective profile from the current HSR measurements. This is expected because of the same facility and nearly matching experimental conditions used for the two experiments. Moreover, the near-wall profiles of ZPG-TBL \citep{willert2018experimental} and APG-TBL \citep{cuvier2017extensive} obtained using time-resolved 2C-2D PIV (TRPIV) have also been included. As the current HSR PIV profiles collapse well with the profiles from the earlier PIV measurements in the respective pressure gradient, this shows sufficient repeatability of the experiments.

\begin{figure}[h]
\begin{center}
  \includegraphics[width=0.95\textwidth]{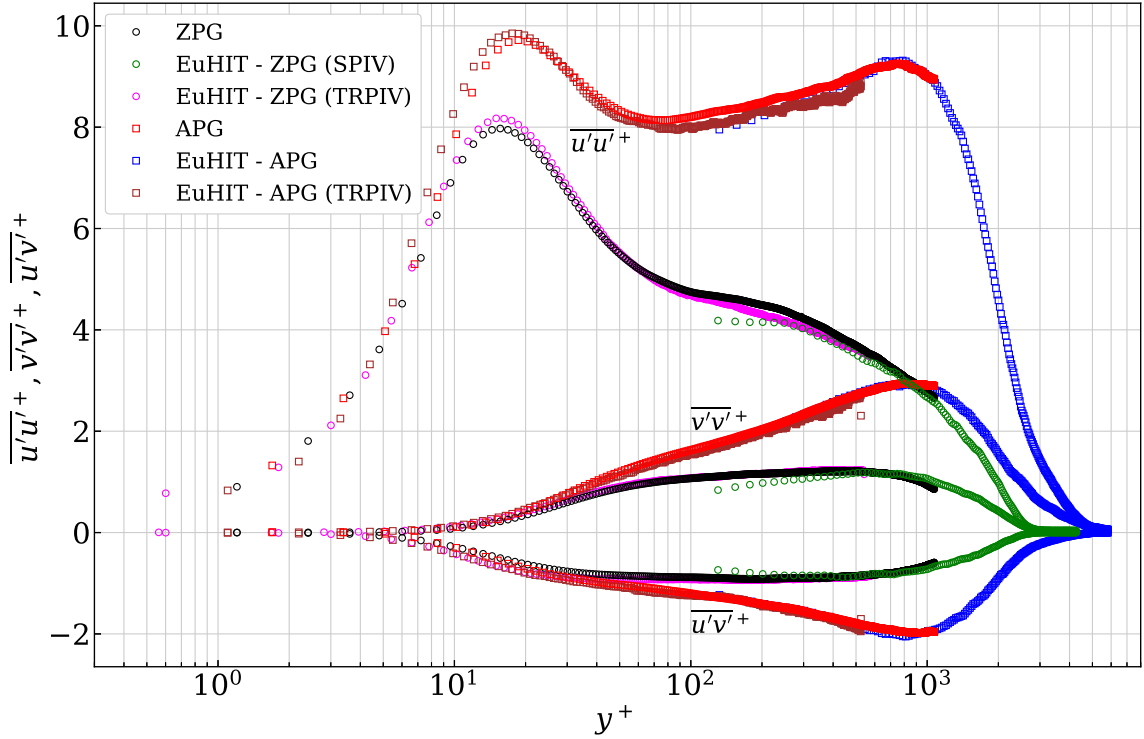}
\caption{Reynolds stress profiles.}
\label{fig:Re_stresses} 
\end{center}
\end{figure}

Reynolds stress profiles scaled in viscous units are shown in Fig. \ref{fig:Re_stresses}. The ZPG-TBL has roughly a $20\%$ lower inner peak in the Reynolds streamwise stress ($\overline{u^\prime u^\prime}$) profile when compared to the APG-TBL and has no outer peak, which is consistent for TBLs in this Reynolds number regime. Whereas in the APG-TBL, the outer peak located at $y^+ \approx 800$ (where $y^+$ denotes the wall viscous units) is almost as strong as the inner peak. The profile of Reynolds wall-normal stress $\overline{v^\prime v^\prime}$ has a less pronounced peak at around $y^+\approx500$ in the ZPG-TBL but is still more than two times weaker than the peak in the APG-TBL, which is also further away from the wall. Reynolds shear stress $\overline{u^\prime v^\prime}$ shows a plateau in the region $y^+\approx 100-300$ in the ZPG-TBL and an outer peak at $y^+\approx900$ in the APG-TBL. This shows that turbulence moves away from the wall towards the outer region when the TBL is under the influence of an APG. Profiles of Reynolds stresses in the outer layer from the EuHIT experiment have also been included for comparison. These profiles deviate by less than $5\%$ from the HSR measurements in the region $y^+ \approx 130-500$ which could be the result of their low-spatial-resolution and more averaging due to larger interrogation windows (IWs) used. Above $y^+=500$, EuHIT profiles are in good agreement with the current HSR measurements. The near-wall profiles of ZPG- and APG-TBL from the TRPIV have also been included. While these profiles are in good agreement with the current HSR measurements for Reynolds wall-normal and shear stresses, they show 2-3\% larger inner peak for Reynolds streamwise stress.

\begin{figure}
\begin{center}
 \includegraphics[width=0.95\textwidth]{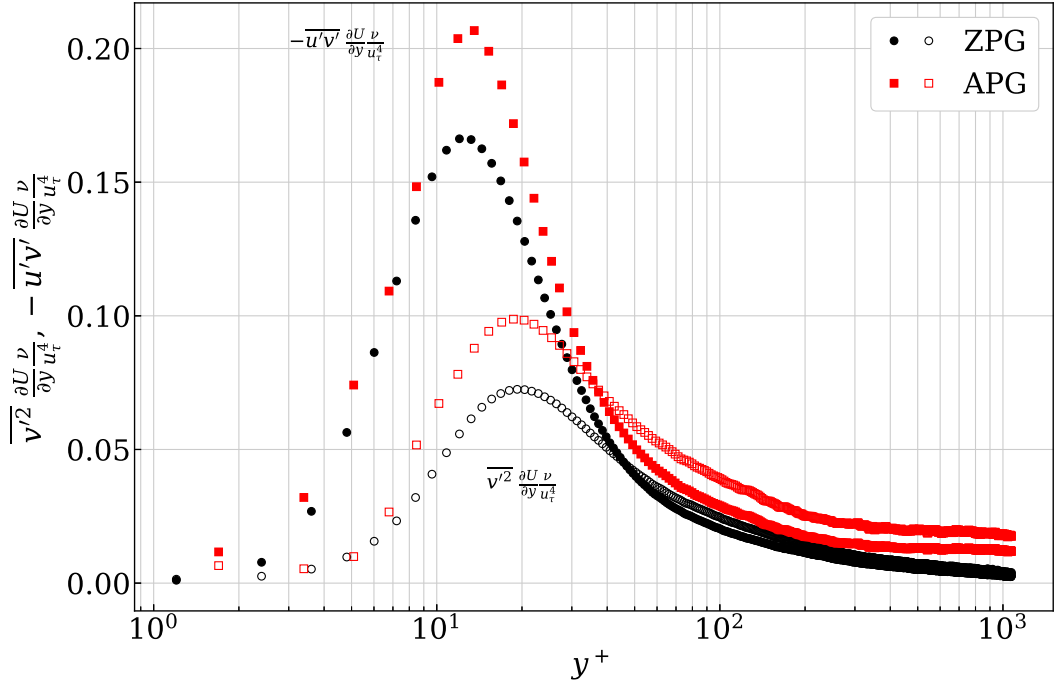}
\caption{Profiles of dominant terms in turbulence production.}
\label{fig:Turbulence_production}  
\end{center}
\end{figure}

The terms that significantly contribute to the turbulence production in the current TBLs are $-\overline{u^\prime v^\prime} \partial U/\partial y$ and $\overline{{v^\prime}^2} \partial U/\partial y$. These terms are present in the equations for $\overline{{u^\prime}^2}$ and $-\overline{u^\prime v^\prime}$ respectively, in the transport equation for the turbulent stress $\overline{u'_i u'_j}$ \citep{skaare1994turbulent}. The profiles of these terms are shown in Fig. \ref{fig:Turbulence_production} where the inner peaks are located at $y^+ \approx12$ and $y^+ \approx18$, respectively. The peak heights of these terms in the APG-TBL are roughly $15\%$ and $30\%$ larger than in the ZPG-TBL. This shows a larger turbulence production in the APG-TBL as compared to the ZPG-TBL in the buffer layer. No significant outer peaks are found in either of these terms. For the APG-TBL, this is in contrast to the findings of \citet{skaare1994turbulent} who found two distinct peaks in the inner and outer regions for each of these terms. They could not resolve the locations of the inner peaks because of their hot-wire data having low-spatial-resolution in the wall-normal direction. However, their outer peaks were observed at locations where stresses are maximum ($y/\delta = 0.45$). The absence of the significant outer peak in the present study is expected because of the mild APG ($\beta \approx 2$) as compared to strong APG in \citet{skaare1994turbulent} ($\beta \approx 21$). In the current measurements, the inner peaks of $-\overline{u^\prime v^\prime} \partial U/\partial y$ are almost two times stronger than the those of $\overline{{v^\prime}^2} \partial U/\partial y$, while both are higher in the APG-TBL than in the ZPG-TBL. 

\subsection{POD Modes}
\label{sec:POD_modes}

\begin{figure}
\begin{center}
\includegraphics[width=0.85\textwidth]{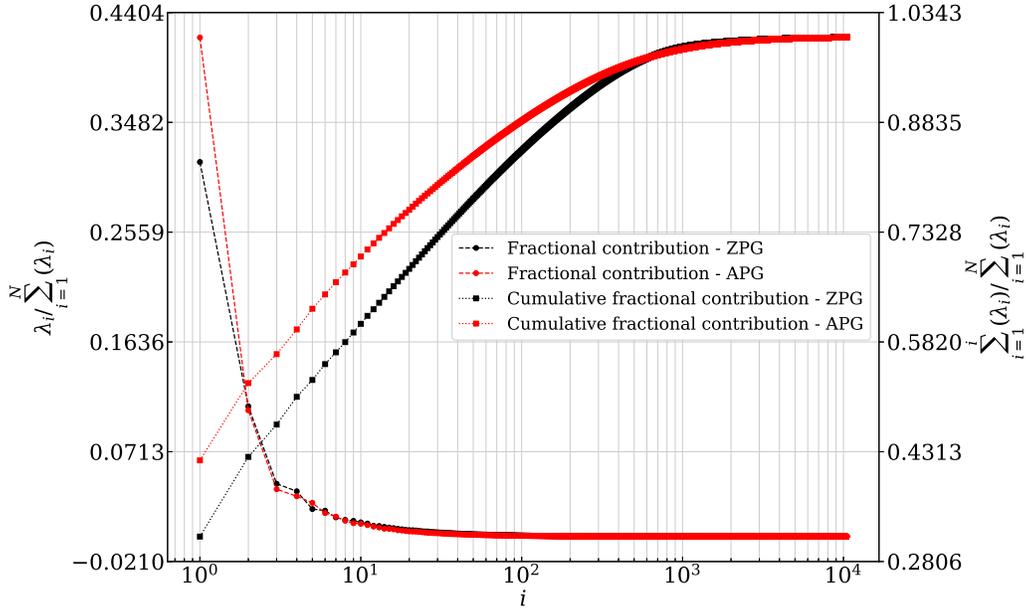}
\caption{Eigenvalues of individual POD modes and their cumulative sum as a fraction of the total energy of all modes.}
\label{fig:Sigma_Z9_A9}    
\end{center}
\end{figure}

As aforementioned, POD was used to analyse the HSR measurements of the ZPG- and APG-TBLs. As the EuHIT TBLs do not cover the inner-layer, POD was not performed on those data-sets to avoid erroneous measurements. Therefore, the results from here onwards will rely only on the HSR measurements. The profiles of eigenvalues corresponding to the individual POD modes and their cumulative sum as a fraction of the total energy are shown in Fig. \ref{fig:Sigma_Z9_A9}. The POD modes are sorted conventionally, such that the contribution to TKE decrease with increasing mode number $i$ i.e. $\lambda_1 \geqslant \lambda_2 \geqslant \; ... \lambda_N \geqslant 0$. The first modes of the ZPG- and APG-TBLs contain $32\%$ and $42\%$ of the total energy in the flow, respectively. For $i\geqslant 2$, the relative contribution of each mode is roughly similar between ZPG- and APG-TBLs. It is also interesting to observe that more than 100 modes are required to obtain 90\% of the energy which indicates the highly stochastic (random) nature of the TBL. 

\begin{figure}[tbph]
\begin{center}
\begin{tabular}{cc}
\includegraphics[trim={0cm 4cm 0cm 0cm},clip,width=\textwidth]{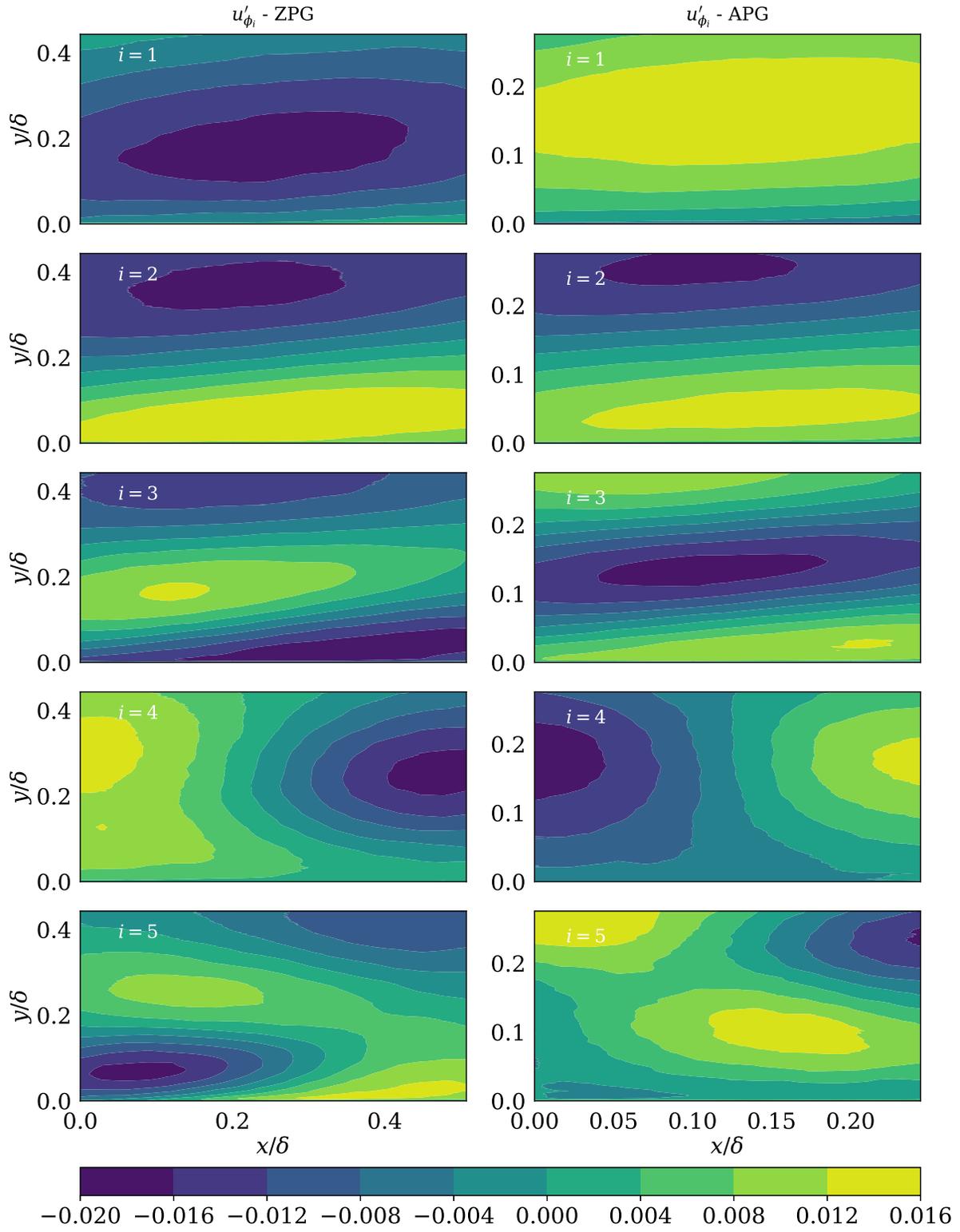} \\
\end{tabular}
\end{center}
\vspace*{-0.2in}\caption{Contour plots of the streamwise components of the first five spatial modes of ZPG-TBL on the left and APG-TBL on the right.}
\label{fig:POD_modes_1to5}
\end{figure}

\begin{figure}[tbph]
\begin{center}
\begin{tabular}{cc}
\includegraphics[trim={0cm 4cm 0cm 0cm},clip,width=\textwidth]{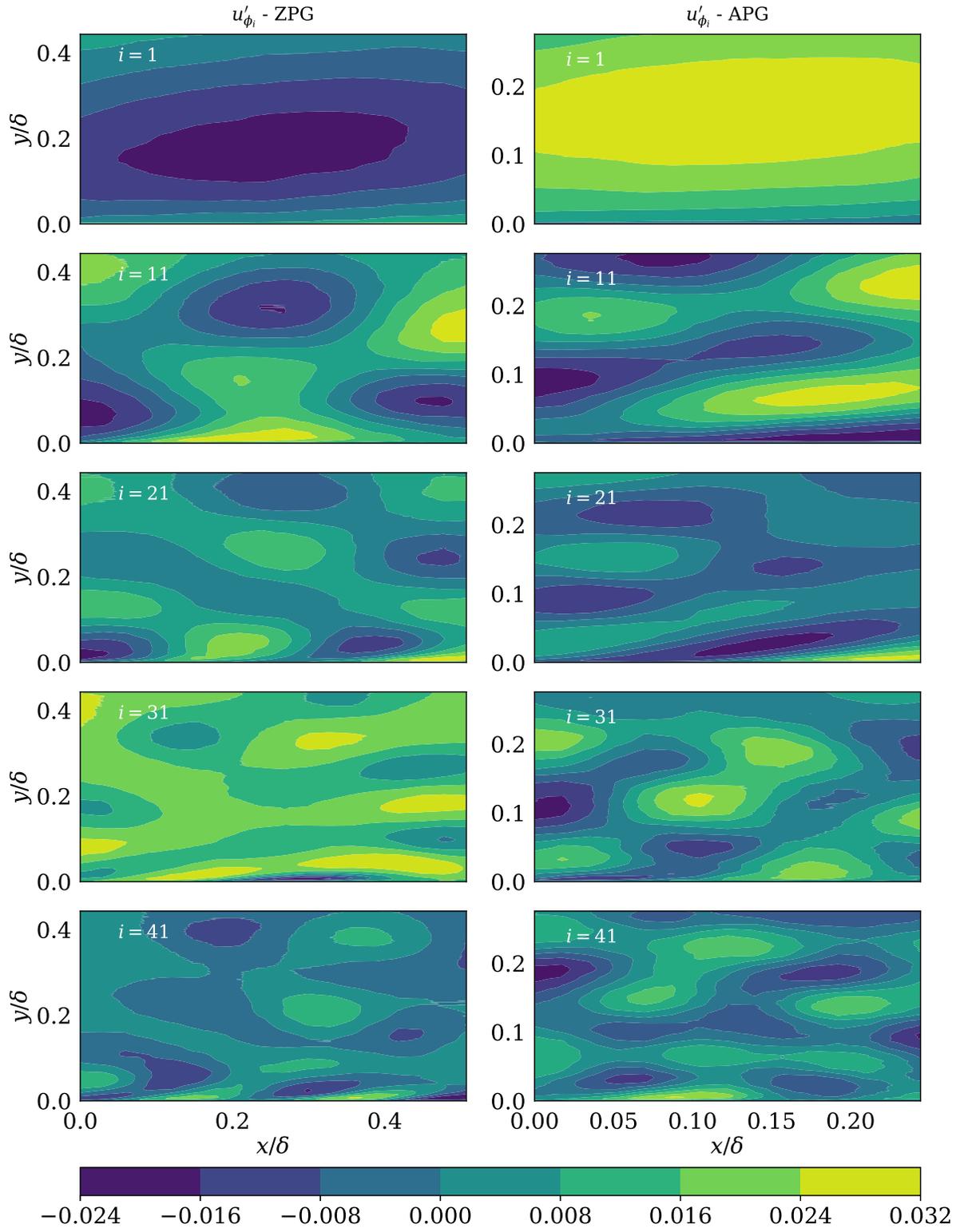} \\
\end{tabular}
\end{center}
\vspace*{-0.2in}\caption{Contour plots of the streamwise components the spatial modes $u'_{\phi_i}$ of ZPG-TBL on the left and APG-TBL on the right where $i=\{1,11,21,31,41\}$.} 
\label{fig:POD_modes_every_10th}
\end{figure}

The contour plots of the streamwise component of the spatial modes $u'_{\phi_i}$ for $i=\{1,2,3,4,5\}$ are shown in figure \ref{fig:POD_modes_1to5} and for $i=\{1,11,21,31,41\}$ in figure \ref{fig:POD_modes_every_10th} for both TBLs. As shown from the two figures, the first mode has the largest scales and hence contributes the most to the TKE. With increasing $i$, the scales become smaller and smaller which is the reason for their reduced contribution to the total TKE.

\begin{figure}[tbph]
\begin{center}
\begin{tabular}{c}
\includegraphics[trim={0cm 0cm 0cm 0cm},clip,width=\textwidth]{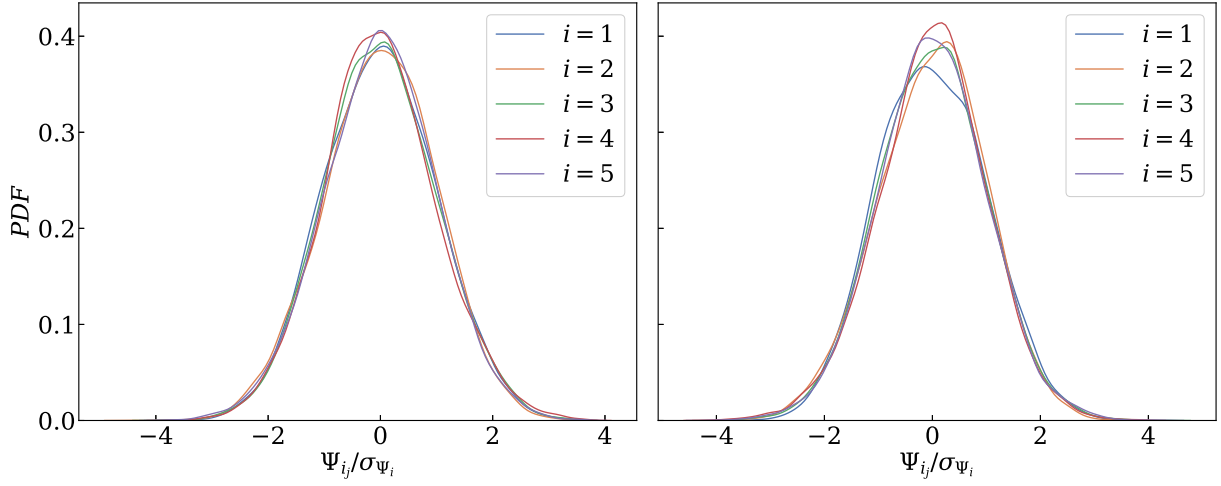} \\
\end{tabular}
\end{center}
\vspace*{-0.2in}\caption{PDFs of temporal coefficients corresponding to the first five most energetic POD modes for (a) ZPG-TBL and (b) APG-TBL. The vertical dotted lines correspond to the $\psi_1 / \sigma_{\psi_1} =\pm1.0$, dashed lines to $\psi_1 / \sigma_{\psi_1} =\pm1.5$ and the solid lines to $\psi_1 / \sigma_{\psi_1} =\pm2.0$.}
\label{fig:PDF_first_5_POD_modes_coefficients}
\end{figure}

Probability density functions (PDFs) of the temporal coefficients corresponding to the first five POD modes for the ZPG- and APG-TBLs are shown in Fig. \ref{fig:PDF_first_5_POD_modes_coefficients}. For each TBL, the first five modes have a similar distribution of their temporal coefficients which appear Gaussian. As all velocity fields (i.e. snapshots) contribute differently to the first and every other POD mode, those fields which have the largest contributions to the large scales of the first mode $\phi_1$ can be characterized as the fields with dominant LSMs. The extent of contribution of each velocity field to the first spatial mode is reflected by the magnitude of its temporal coefficient corresponding to $\phi_1$. This is an enhanced definition of the basis of the classification described in section \ref{sec:classification}. The threshold $K$ in Eq. \ref{eq:classification_0} is selected based on the distribution of $\psi_1$. Given that the temporal coefficients of the first POD mode have a Euclidean norm of 1 (similar to results of \citet{guemesexperimental}), $K=1$ is chosen for the classification of LSMs into high and low-momentum events. A sensitivity analysis of the second-order turbulent statistics without LSMs using the classification based on the different values of the threshold factor $K=\{1.0,1.5,2.0\}$ is presented in section \ref{sec:sensitivity_analysis}.

\begin{figure}[tbph]
\begin{center}
\begin{tabular}{c}
\includegraphics[trim={0cm 0cm 0cm 0cm},clip,width=\textwidth]{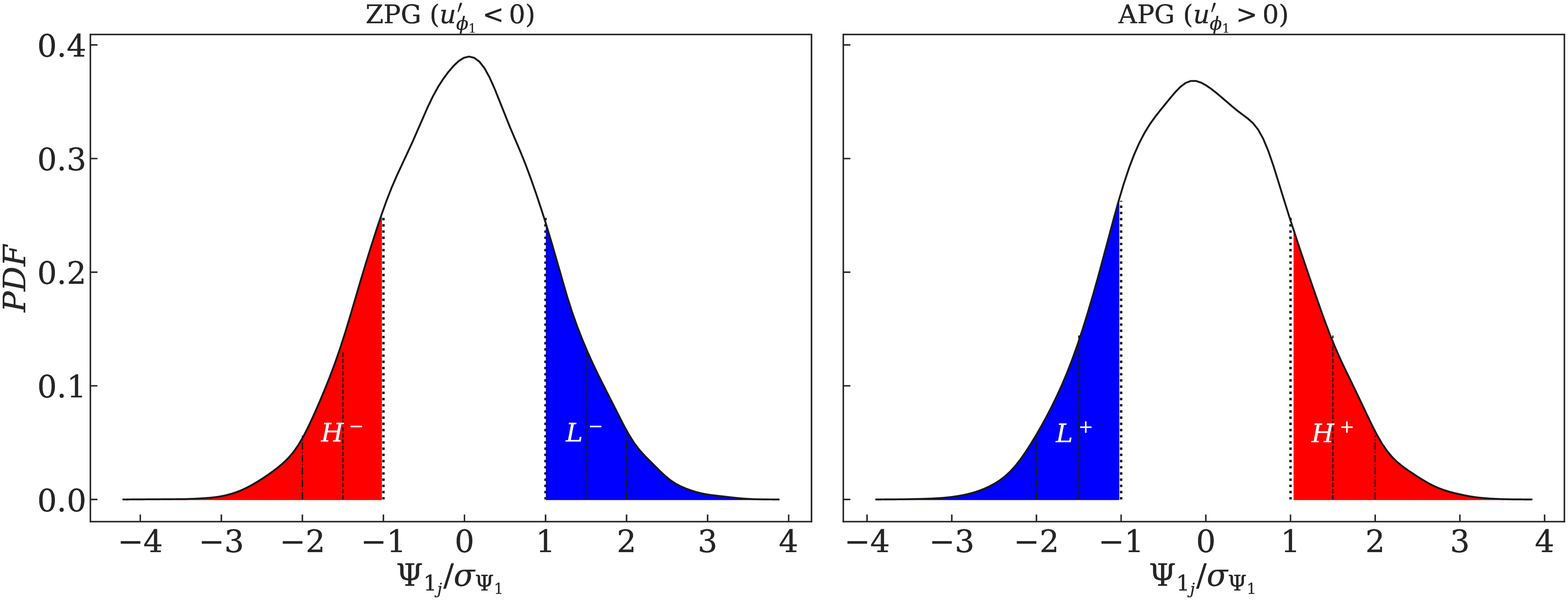} \\
\end{tabular}
\end{center}
\vspace*{-0.2in}\caption{Classification of LSMs in to high- and low-momentum events depicted on the PDFs of temporal coefficients corresponding to the first POD modes. The vertical dotted lines correspond to the $\psi_1 / \sigma_{\psi_1} =\pm1.0$, dashed lines to $\psi_1 / \sigma_{\psi_1} =\pm1.5$ and the dash-dotted lines to $\psi_1 / \sigma_{\psi_1} =\pm2.0$.}
\label{fig:PDF_first_mode_coefficients_with_classification}
\end{figure}

To investigate the effect of high- and low-momentum LSMs on turbulent statistics, the flow fields $F$ of each TBL is divided into three parts: (i) high-momentum LSMs ($H$), (ii) low-momentum LSMs ($L$) and (iii) SSMs. Further classification of the fields with LSMs into those dominated by high- or low-momentum events is based on the signs of $\psi_{1_j}$ and $u'_{\phi_1}$. As $u'_{\phi_1}<0$ for the ZPG-TBL, the velocity fields of the ZPG-TBL with high- and low-momentum events are $H^-$ and $L^-$, respectively. In contrast, as $u'_{\phi_1}>0$ for the APG-TBL, the velocity fields of the APG-TBL with high- and low-momentum events are $H^+$ and $L^+$, respectively. The distributions of the temporal coefficients of velocity fields with LSMs for both TBLs are shown in figure \ref{fig:PDF_first_mode_coefficients_with_classification} where the red regions correspond to high-momentum events and the blue regions correspond to low-momentum events. These regions are bounded by dotted lines of $\psi_{1_j}=\pm \sigma_{\psi_1}$ and solid lines of $PDF(|\psi_{1_j}| > \sigma_{\psi_1})$. The dashed lines and dash-dotted lines represent $K=1.5$ and $K=2.0$ from figure \ref{eq:classification_0}, respectively.

\newpage
\subsection{Conditional Averaging of the Turbulent Statistics}
\label{sec:conditionally_averaged_stats}

The conditional averaging of turbulent statistics is performed on the sub-fields $H$ and $L$ as presented in table \ref{tab:conditional_averaging}. The $-$ and $+$ superscripts differentiate between the nature of the first POD mode of the TBLs (see sections \ref{sec:classification} and \ref{sec:POD_modes}). The subscripts $h$ and $l$ represent the conditionally averaged statistics from the $H$ and $L$ sub-fields, respectively. This notation will be followed in the profiles of the conditionally averaged turbulent statistics. 

\begin{table}[ht]
\begin{center}
\caption{The nomenclature of the conditionally averaged turbulent statistics.}
\renewcommand{\arraystretch}{1.5}
\begin{tabular}{wl{1.3cm}wl{2cm}wl{2cm}|wl{1.3cm}wl{2cm}wl{2cm}}
\hline \hline\noalign{\medskip}
 \multicolumn{3}{c}{High-momentum} & \multicolumn{3}{c}{Low-momentum }  \\
Property & ZPG &APG & Property & ZPG & APG\\
\hline
$U_h$     & $\overline{u[H^-]}$   & $\overline{u[H^+]}$  &$U_l$      & $\overline{u[L^-]}$     & $\overline{u[L^+]}$ \\
$\overline{u'u'}_h$  & $\overline{({u'[H^-]})^2}$ & $\overline{({u'[H^+]})^2}$  &$\overline{u'u'}_l$  & $\overline{({u'[L^-]})^2}$   & $\overline{({u'[L^+]})^2}$ \\
$\overline{v'v'}_h$  & $\overline{({v'[H^-]})^2}$    & $\overline{({v'[H^+]})^2}$  &$\overline{v'v'}_l$ & $\overline{({v'[L^-]})^2}$    & $\overline{({v'[L^+]})^2}$ \\
$\overline{u'v'}_h$  & $\overline{u'[H^-]v'[H^-]}$  & $\overline{u'[H^+]v'[H^+]}$  &$\overline{u'v'}_l$ & $\overline{u'[L^-]v'[L^-]}$  & $\overline{u'[L^+]v'[L^+]}$ \\
\hline
\label{tab:conditional_averaging} 
\end{tabular}
\end{center}
\end{table}

\begin{figure}[tbph]
\begin{center}
\begin{tabular}{cc}
\includegraphics[trim={2.1cm 0cm 0cm 0cm},clip,width=0.43\textwidth]{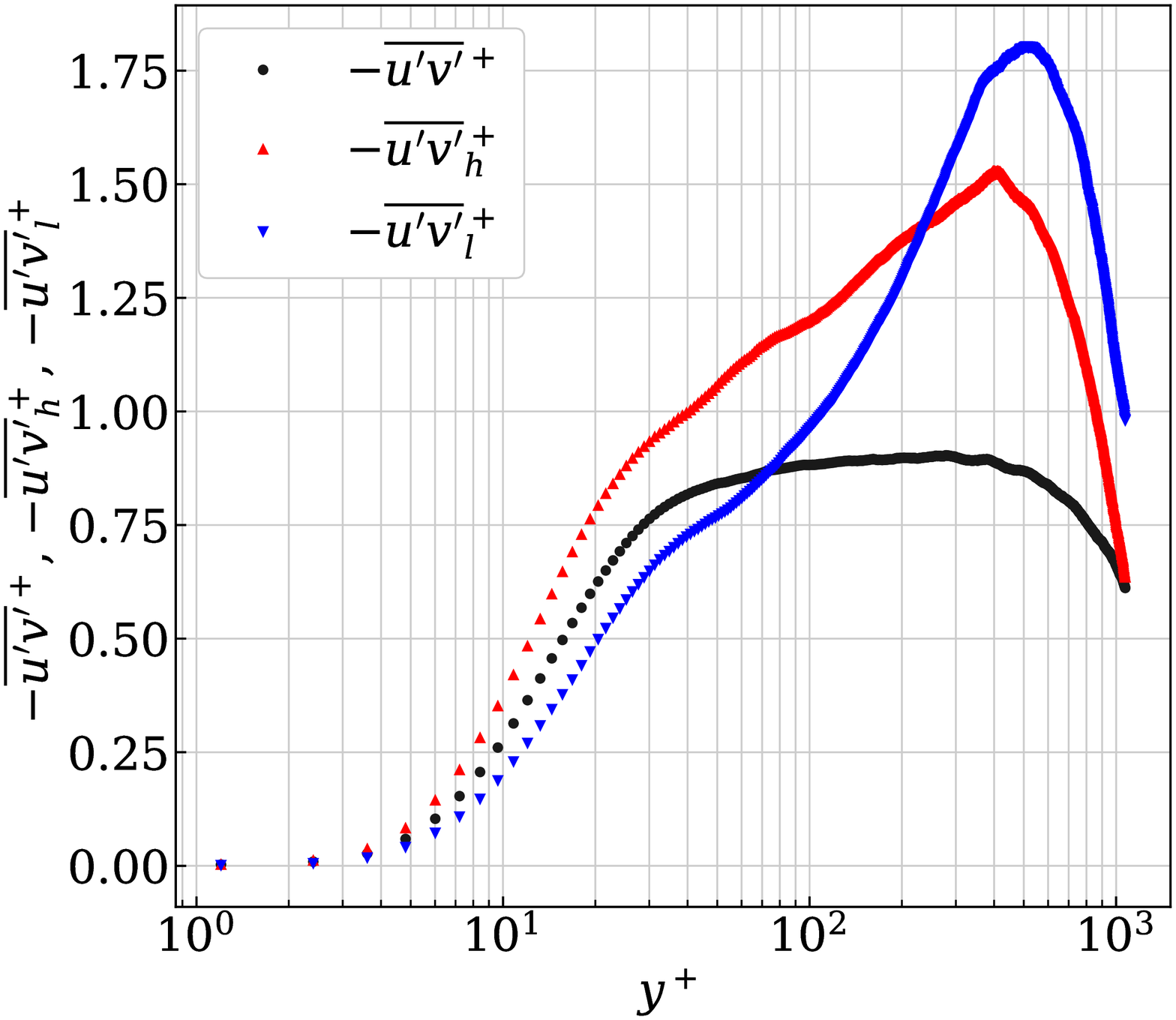} &
\includegraphics[trim={2.2cm 0cm 0cm 0cm},clip,width=0.43\textwidth]{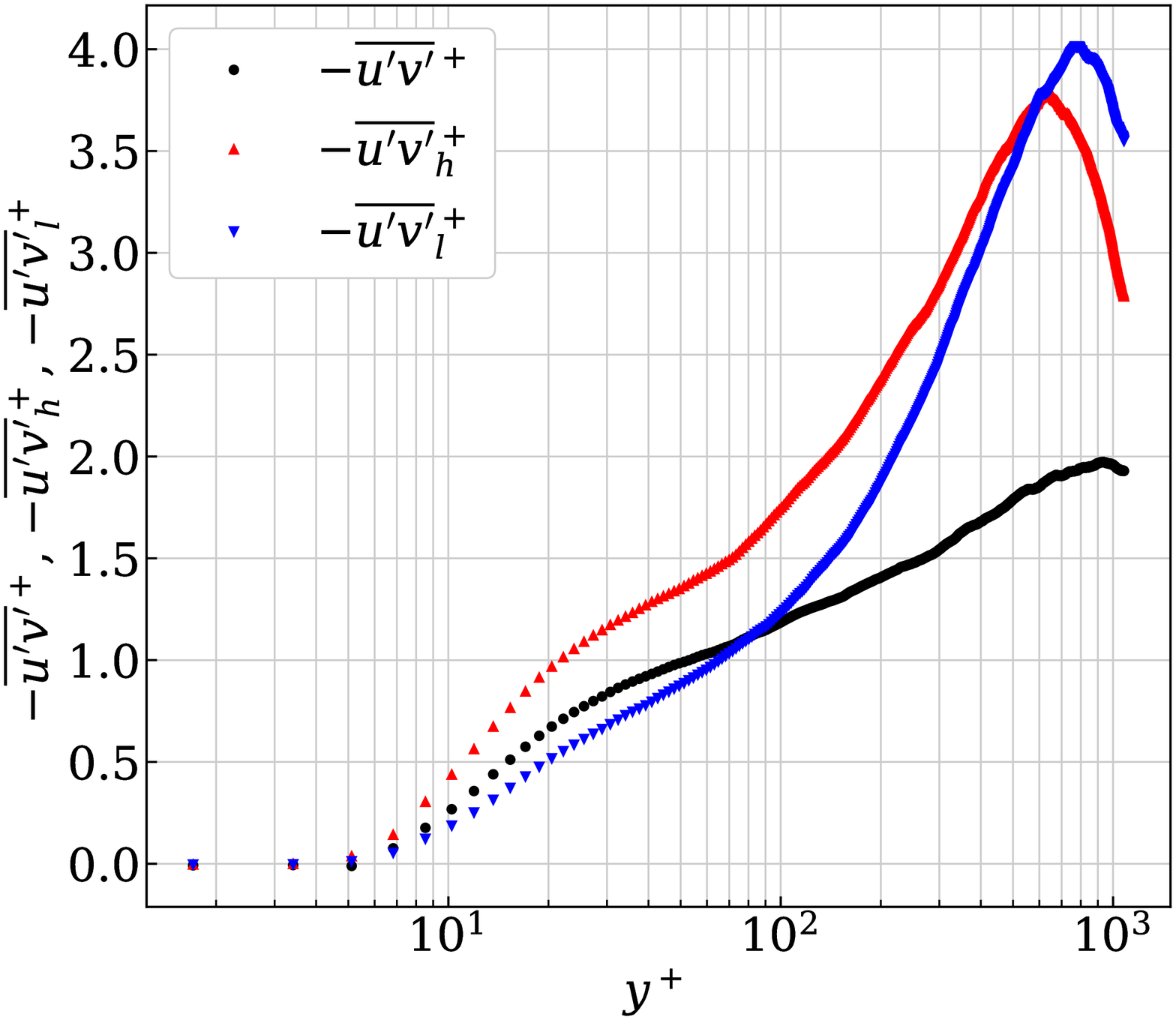}  \\
(g) & (h) \\
\includegraphics[trim={2.5cm 0cm 0cm 0cm},clip,width=0.43\textwidth]{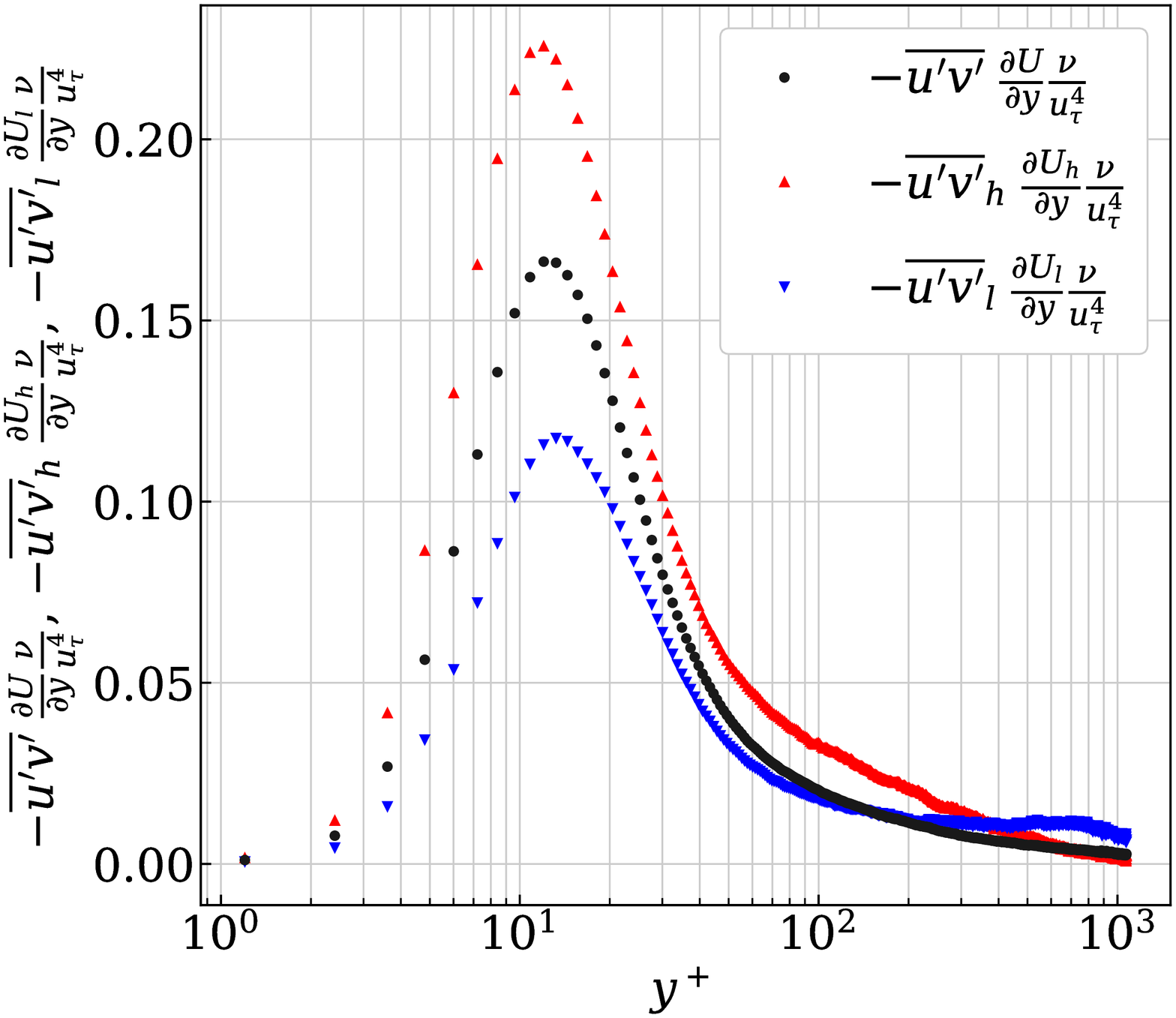} &
\includegraphics[trim={2.5cm 0cm 0cm 0cm},clip,width=0.43\textwidth]{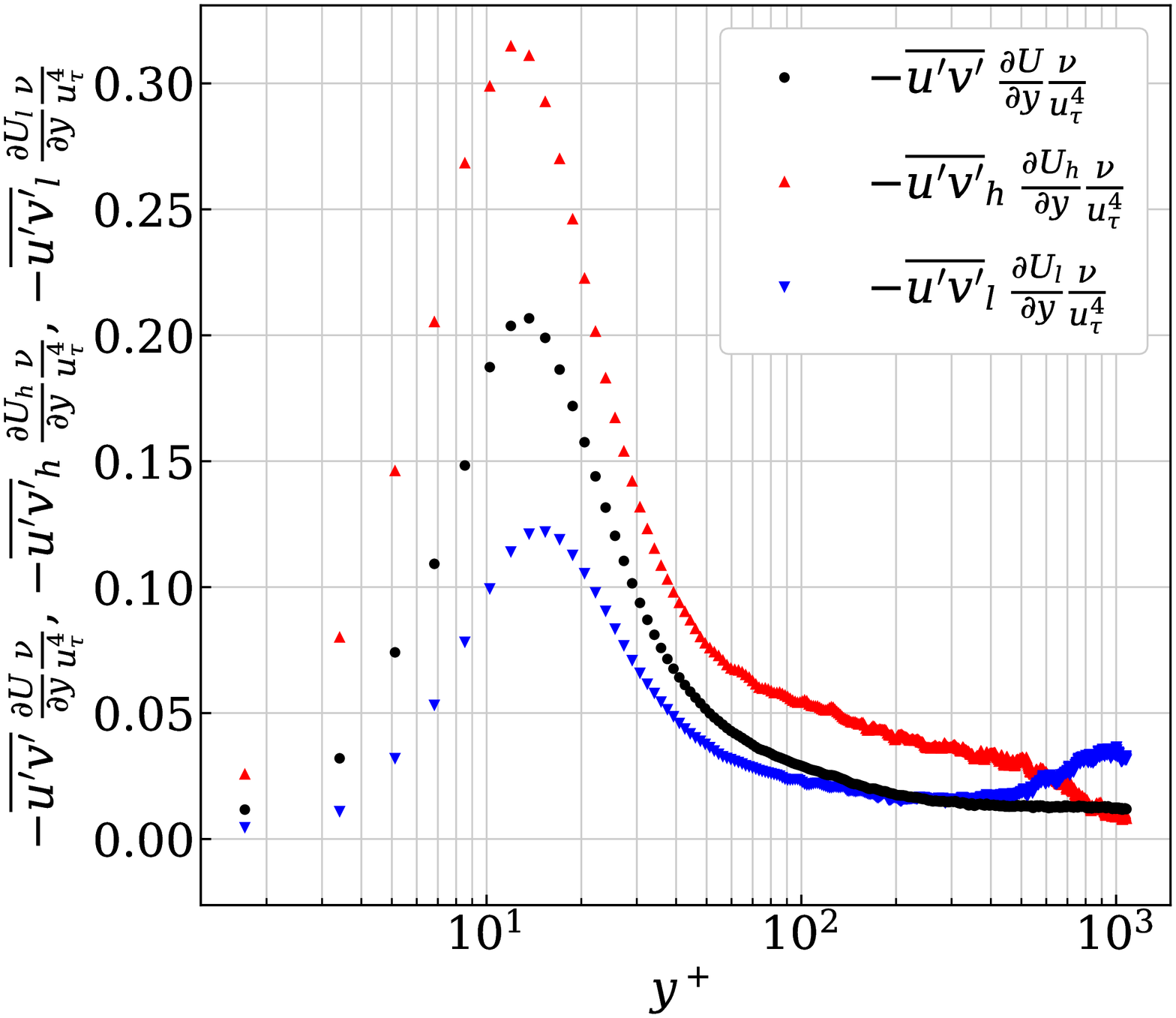}  \\
(i) & (j) \\
\end{tabular}
\end{center}
\caption{Mean streamwise velocity, Reynolds streamwise, wall-normal and shear stresses and turbulence production term showing the effect of high- and low-momentum LSMs in ZPG-TBL (a),(c),(e),(g),(i), and APG-TBL (b),(d),(f),(h),(j), respectively. 
\label{fig:conditionally_ave_stats}}
\end{figure}

The profiles of these statistics, along with the dominant term in turbulence production $-\overline{u'v'}\frac{\partial U}{\partial y}$ scaled in viscous units are shown in figure  \ref{fig:conditionally_ave_stats} for the ZPG- and APG-TBLs. The profiles from the original ensemble have also been included for comparison. As shown in the mean velocity profiles of the ZPG-TBL (figure \ref{fig:conditionally_ave_stats}(a)), high-momentum events have a higher mean velocity when compared to the original ensemble, while the opposite is true for the low-momentum events. A similar trend is observed in the APG-TBL (figure \ref{fig:conditionally_ave_stats}(b)), but profiles of high- and low-momentum motions are farther from the original ensemble in the outer region when compared to the ZPG-TBL which shows that the activity of the LSMs is enhanced by the APG in the outer layer. 

For the Reynolds streamwise stress in the ZPG-TBL (figure \ref{fig:conditionally_ave_stats}(c)), the high-momentum events contribute more near the wall and in the log region when compared to the low-momentum events while the opposite is the case in the wake region. Near the wall, high-momentum events have positive contributions and low-momentum events have negative contributions to Reynolds streamwise stress. This is in agreement with observations of \citet{guemesexperimental}, who performed similar investigations in a ZPG-TBL. The near-wall peak in $\overline{u'u'}_h$ is about $22\%$ more pronounced compared to the original ensemble. Both high- and low-momentum events produce an outer peak that is almost as strong as the inner peak in the original ensemble. 

Similarly, high-momentum events contribute more to the Reynolds wall-normal stresses than low-momentum from the wall to the end of the log-layer and the opposite is the case in the wake region (see figure \ref{fig:conditionally_ave_stats}(e)). The same near-wall distribution is also found in Reynolds shear stress (see figure \ref{fig:conditionally_ave_stats}(g)). The Reynolds streamwise and shear stress profiles for high- and low-momentum events are up to two times stronger in the wake region compared to the corresponding profiles from the original ensembles. This leads to the conclusion that the LSMs have significantly strong activity in the outer region, especially in the streamwise velocity fluctuations. Another clear observation in the Reynolds stresses is that the cross-over point of the high- and low-momentum profiles happens to be at values higher than the original ensemble which is contradictory to the findings of \citet{guemesexperimental}. Moreover, \cite{guemesexperimental} found that the cross-over points in all Reynolds stresses are at nearly the same wall-normal location, which is not the case in the present study. This is probably because the HSR data in this study does not cover the whole boundary layer, which affects the output of POD and hence the classification such that the cross-over point falls at different $y$ location for different Reynolds stresses. 

In the APG-TBL case, the contributions of the high- and low-momentum LSMs to Reynolds stresses (see figure \ref{fig:conditionally_ave_stats}-\{(d),(f),(h)\}) show similar trends as in the ZPG-TBL, but their outer peaks are slightly more distant from the original ensemble. This shows the further activation of LSMs in the presence of an APG. Another observable difference is that the cross-over point of the LSM profiles in the APG-TBL moves further from the wall than in the ZPG-TBL. 

The profiles of $-\overline{u'v'}\frac{\partial U}{\partial y}$ show that high- and low-momentum LSMs have inner peaks which are about 25\% stronger and weaker than the original ensemble respectively (see figure \ref{fig:conditionally_ave_stats}(i)). The cross-over point in these profiles is also above the original ensemble and is located near the wall-normal location of the cross-over point in the Reynolds shear stress profiles. These changes in the turbulence production near the wall are in agreement with the results of \citet{hutchins2007evidence}, who suggest that the LSMs of the log layer have footprints in the near-wall region. In the APG-TBL (see figure \ref{fig:conditionally_ave_stats}(j)), this spread between the high- and low-momentum LSMs is about $50\%$ larger than in the ZPG-TBL, and their cross-over point is also farther in wall-normal direction. This shows that the TKE production due to the high- and low-momentum events is enhanced and reduced, respectively, by the presence of an APG.

\begin{figure}[tbph]
\begin{center}
\begin{tabular}{cc}
\includegraphics[trim={2cm 0cm 0cm 0cm},clip,width=0.43\textwidth]{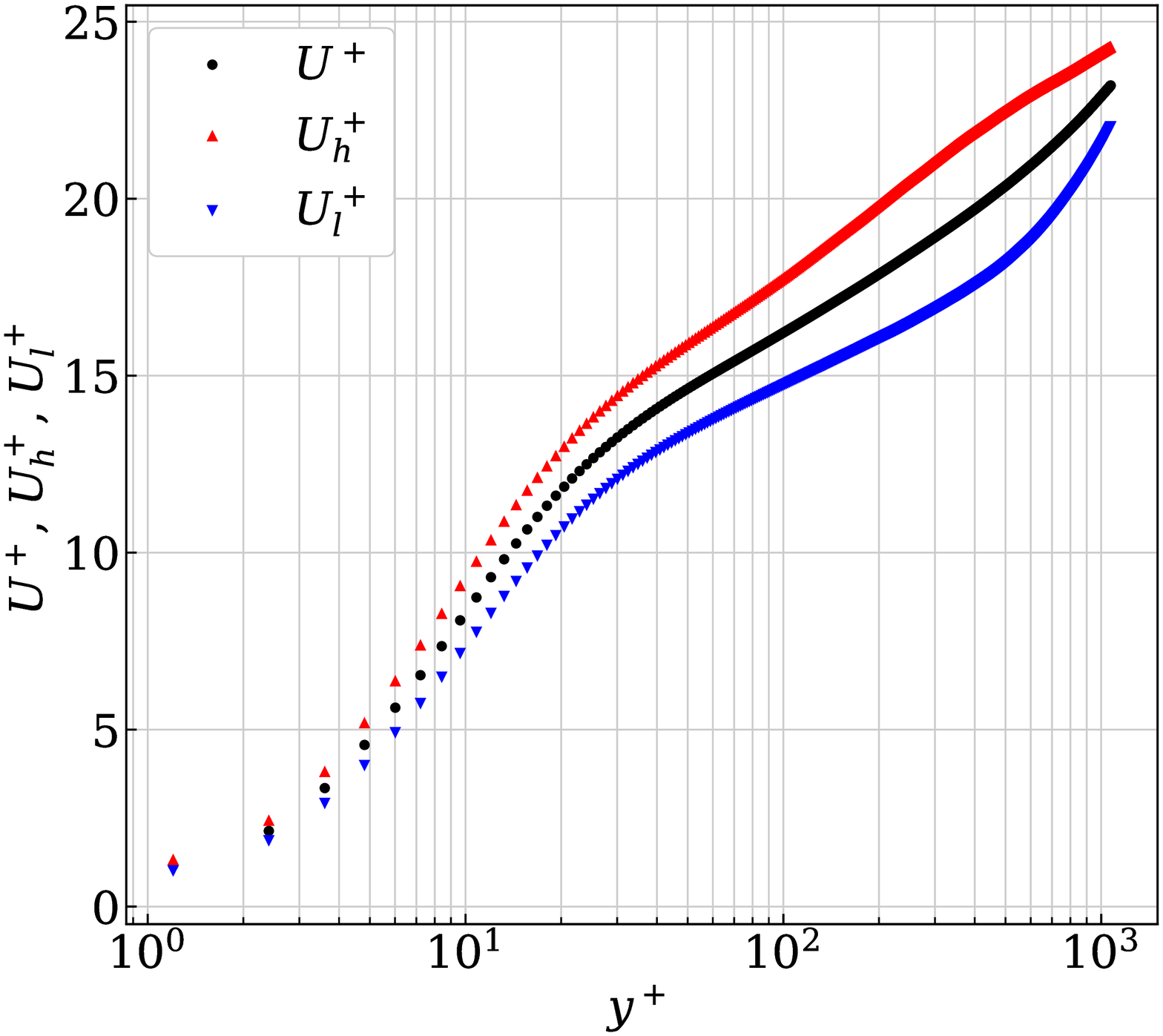} &
\includegraphics[trim={2cm 0cm 0cm 0cm},clip,width=0.43\textwidth]{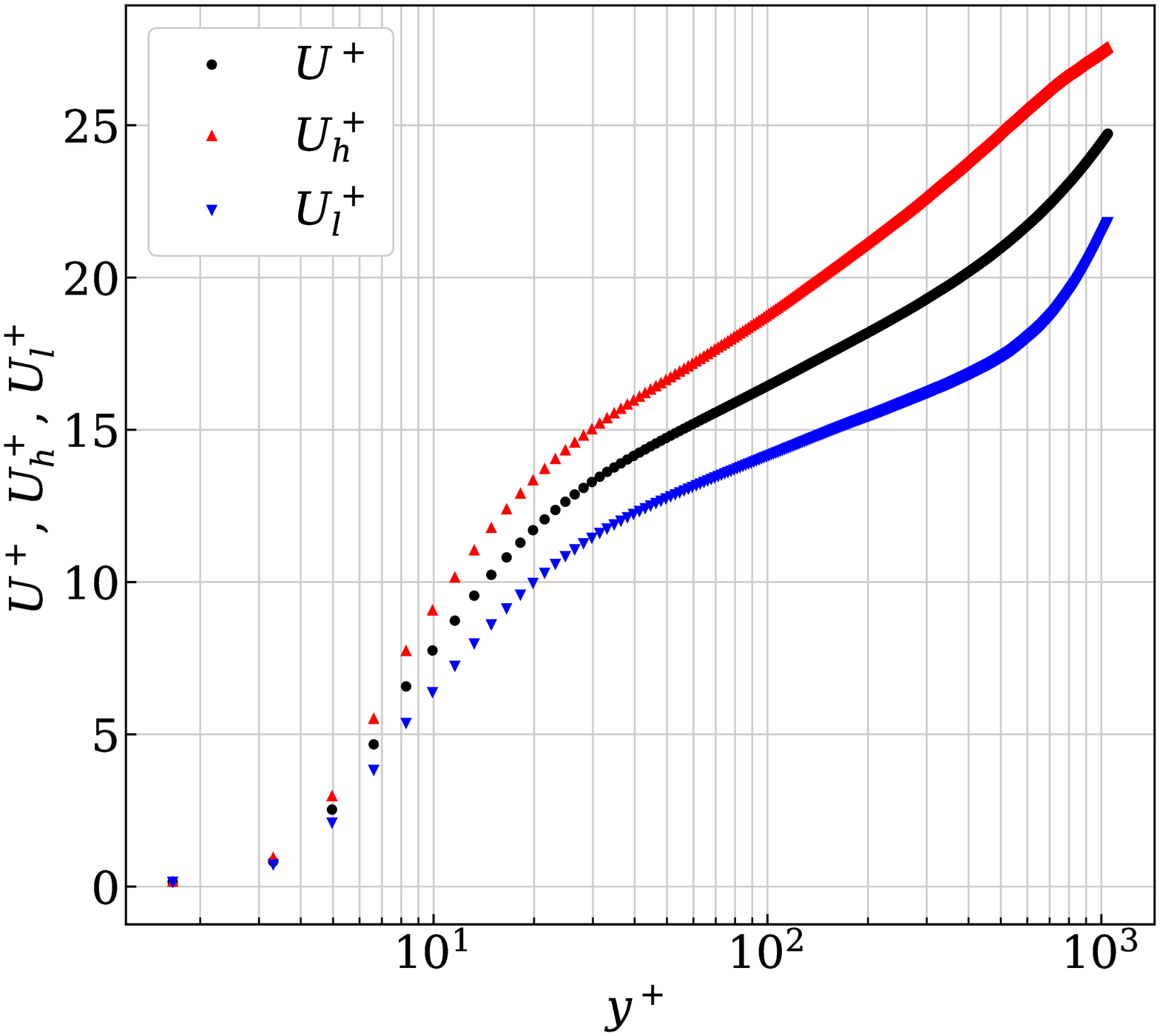} \\
(a) & (b)\\
\includegraphics[trim={2.1cm 0cm 0cm 0cm},clip,width=0.43\textwidth]{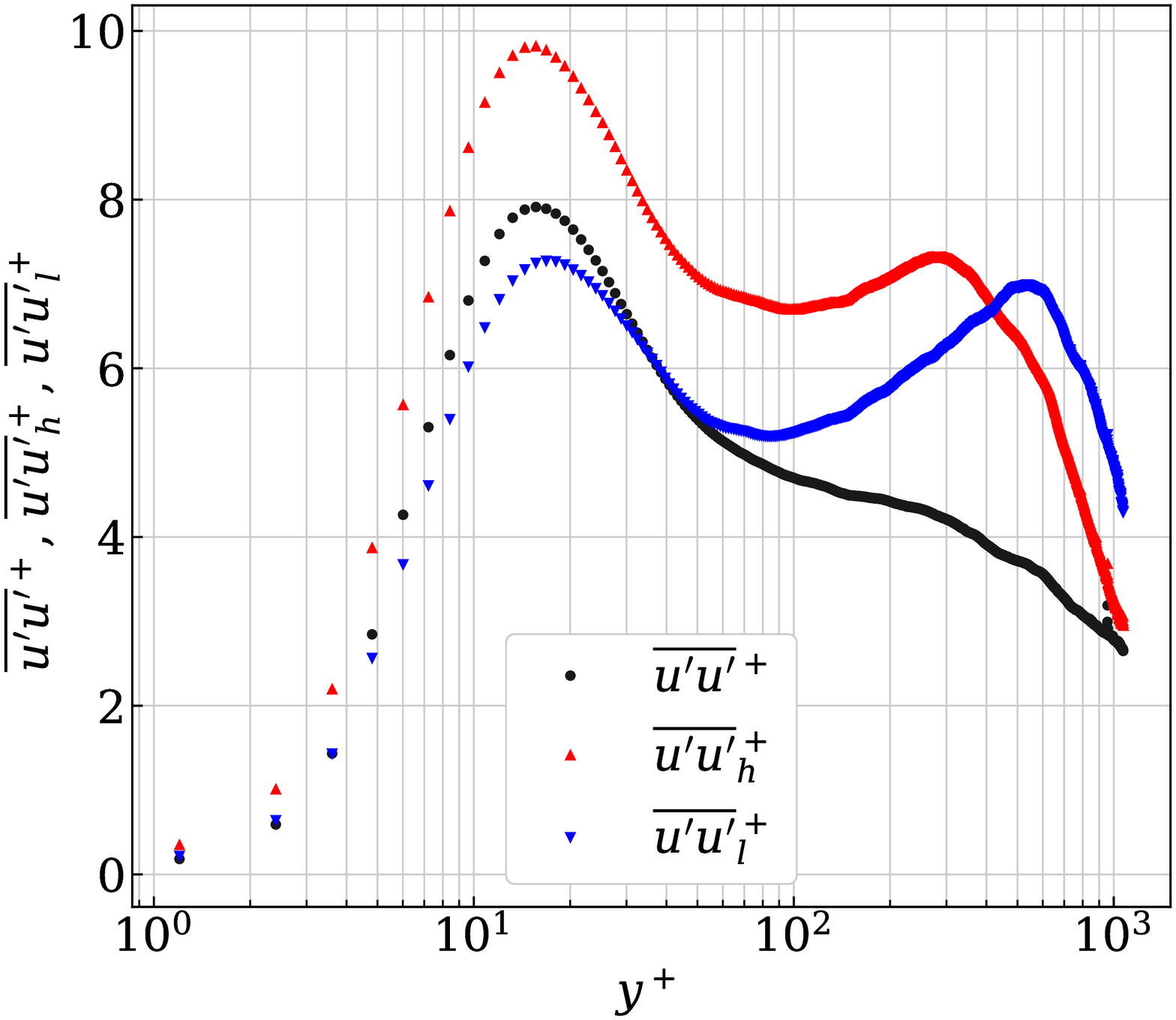} &
\includegraphics[trim={2.2cm 0cm 0cm 0cm},clip,width=0.43\textwidth]{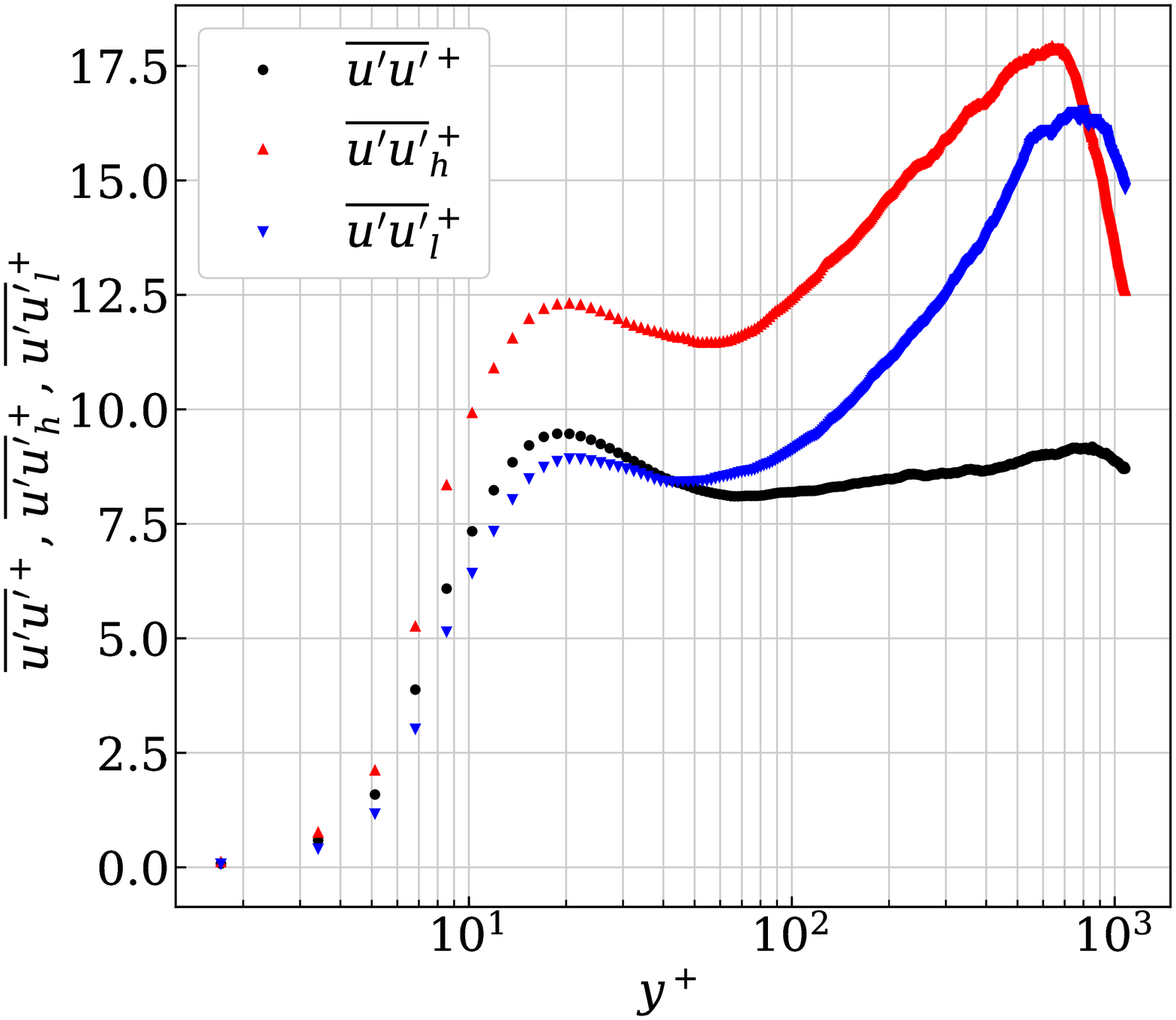}  \\
(c) & (d) \\
\includegraphics[trim={2.1cm 0cm 0cm 0cm},clip,width=0.43\textwidth]{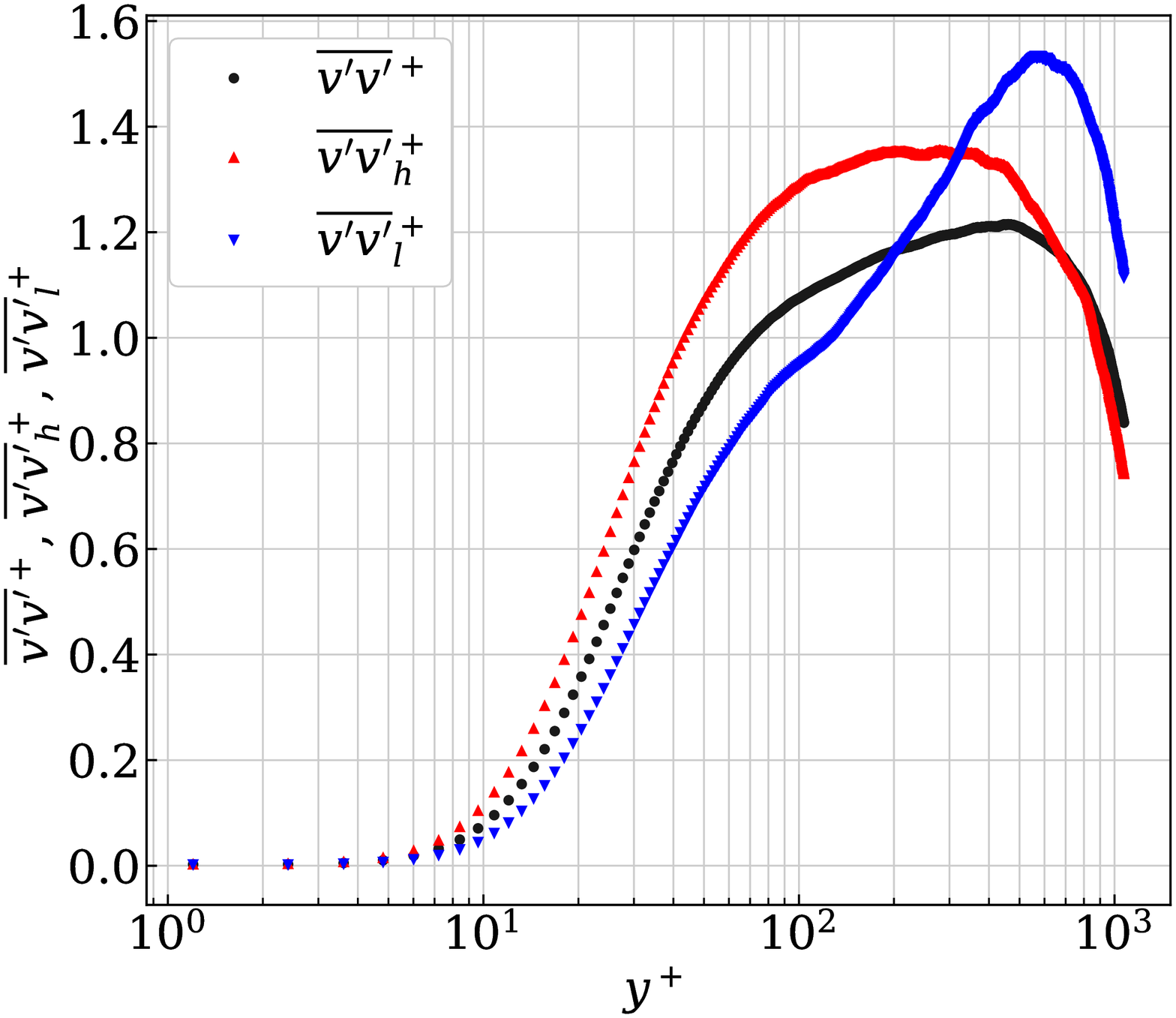} &
\includegraphics[trim={2.2cm 0cm 0cm 0cm},clip,width=0.43\textwidth]{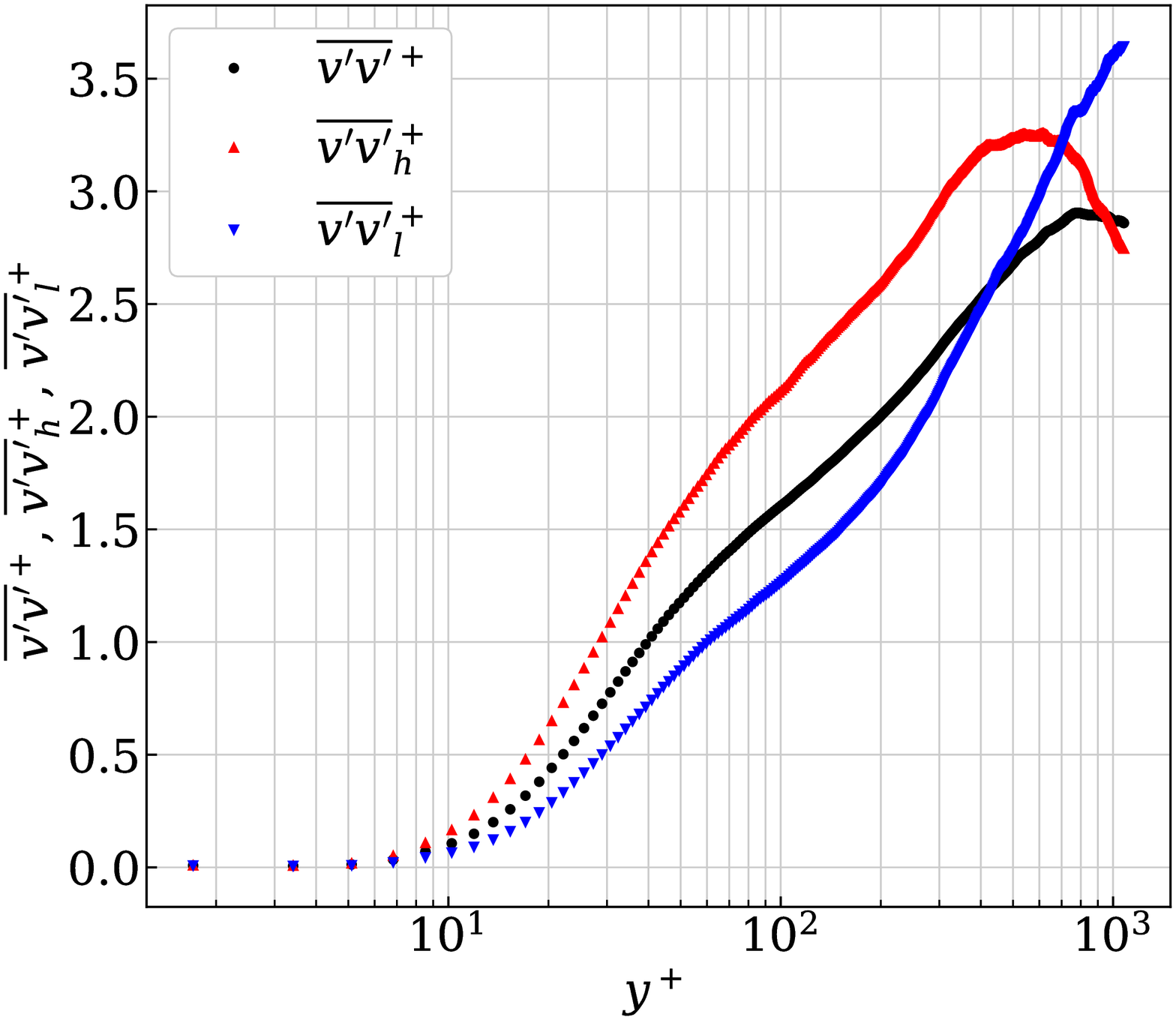}  \\
(e) & (f) \\
\end{tabular}
\end{center}
\phantomcaption{See caption on the next page.}
\end{figure}

\subsection{Sensitivity analysis}
\label{sec:sensitivity_analysis}

As described in section \ref{sec:classification}, the flow-field is separated into large- and small-scale motions based on a threshold value of the temporal coefficient of the first ({\em i.e.} the most energetic) POD mode,  $\psi_{1_j}$. As  $\psi_1$ has a unitary Euclidean norm, the threshold factor $K$ is taken as $1$. It is, however, worthwhile to investigate the effect of larger values of $K$ (i.e. $1.5$ and $2$) on the conditionally averaged turbulent statistics to find out what portion of LSMs will be subtracted if the larger values of $K$ are used. Because of the Gaussian nature of the distribution of $\psi_1$ as shown in figure \ref{fig:PDF_first_5_POD_modes_coefficients}, fewer and fewer number of velocity fields match the criterion $|\psi_1|>K\sigma_{\psi_1}$ for values of $K$ increasing from $1.0$ to $2.0$. Table \ref{tab:percent_fieds_for_different_K} lists the percentage values of the number of the velocity fields matching this criterion for $K=\{1.0,1.5,2.0\}$. As evident, the effect of a mild APG on these statistics is negligible as these statistics are very similar for both TBLs. 

\begin{table}[ht]
\begin{center}
\caption{Percentage of velocity fields with $|\psi_1|>K\sigma_{\psi_1}$ for $K=\{1.0,1.5,2.0\}$.}
\renewcommand{\arraystretch}{1.3}
\begin{tabular}{wc{1.5cm}wc{2cm}wc{2cm}}
\hline \hline\noalign{\medskip}
 & ZPG & APG \\
$K$ & (\%) & (\%) \\

\hline
1.0 & 32.7 & 33.0 \\
1.5 & 13.6 & 13.2 \\
2.0 & 4.2 & 4.0 \\
\hline
\label{tab:percent_fieds_for_different_K} 
\end{tabular}
\end{center}
\end{table}

\begin{figure}[tbph]
\begin{center}
\begin{tabular}{cc}
\includegraphics[trim={1cm 0cm 0cm 0cm},clip,width=0.43\textwidth,height=2.5in]{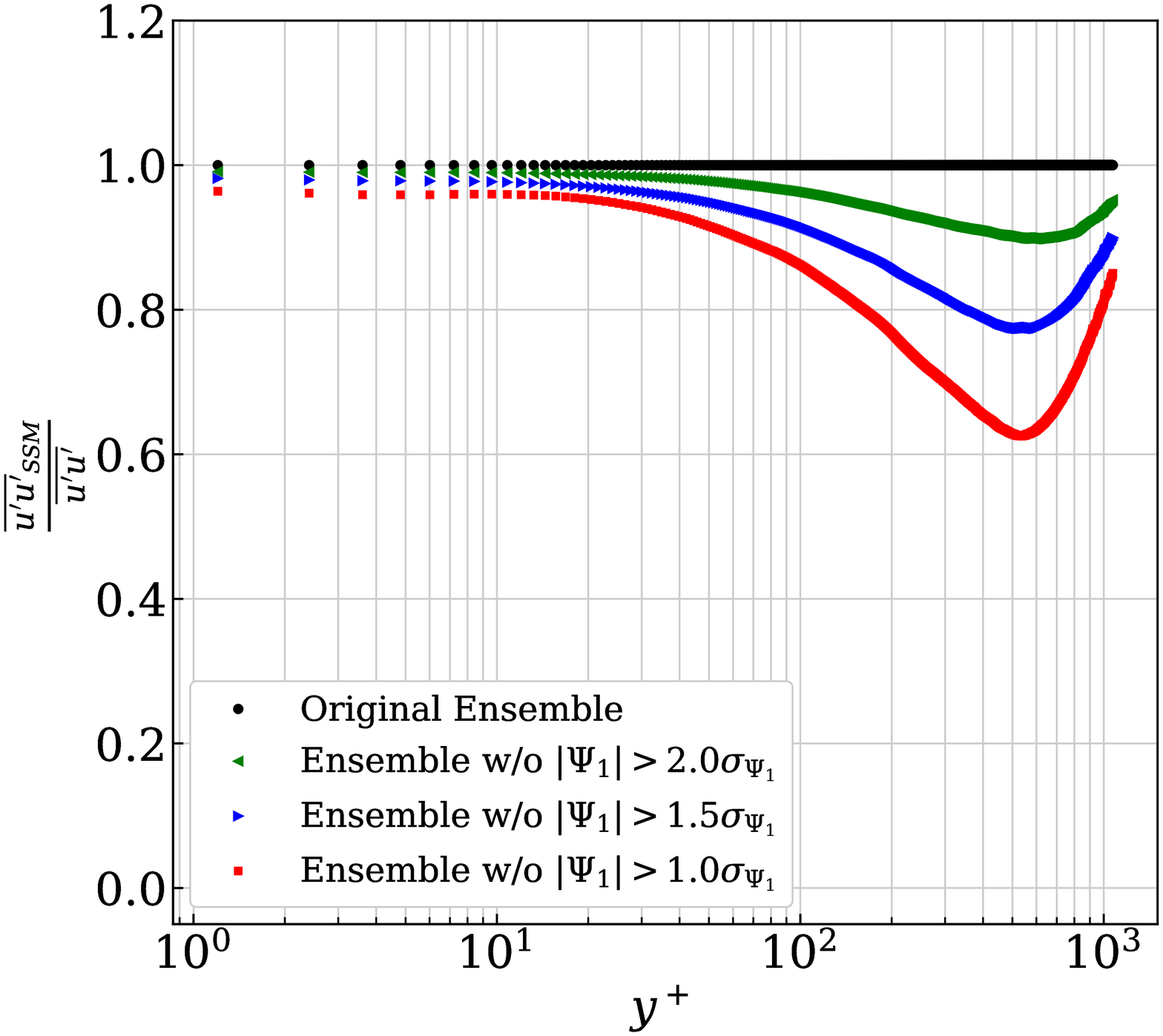} & \includegraphics[trim={2.55cm 0cm 0cm 0cm},clip,width=0.4\textwidth,height=2.5in]{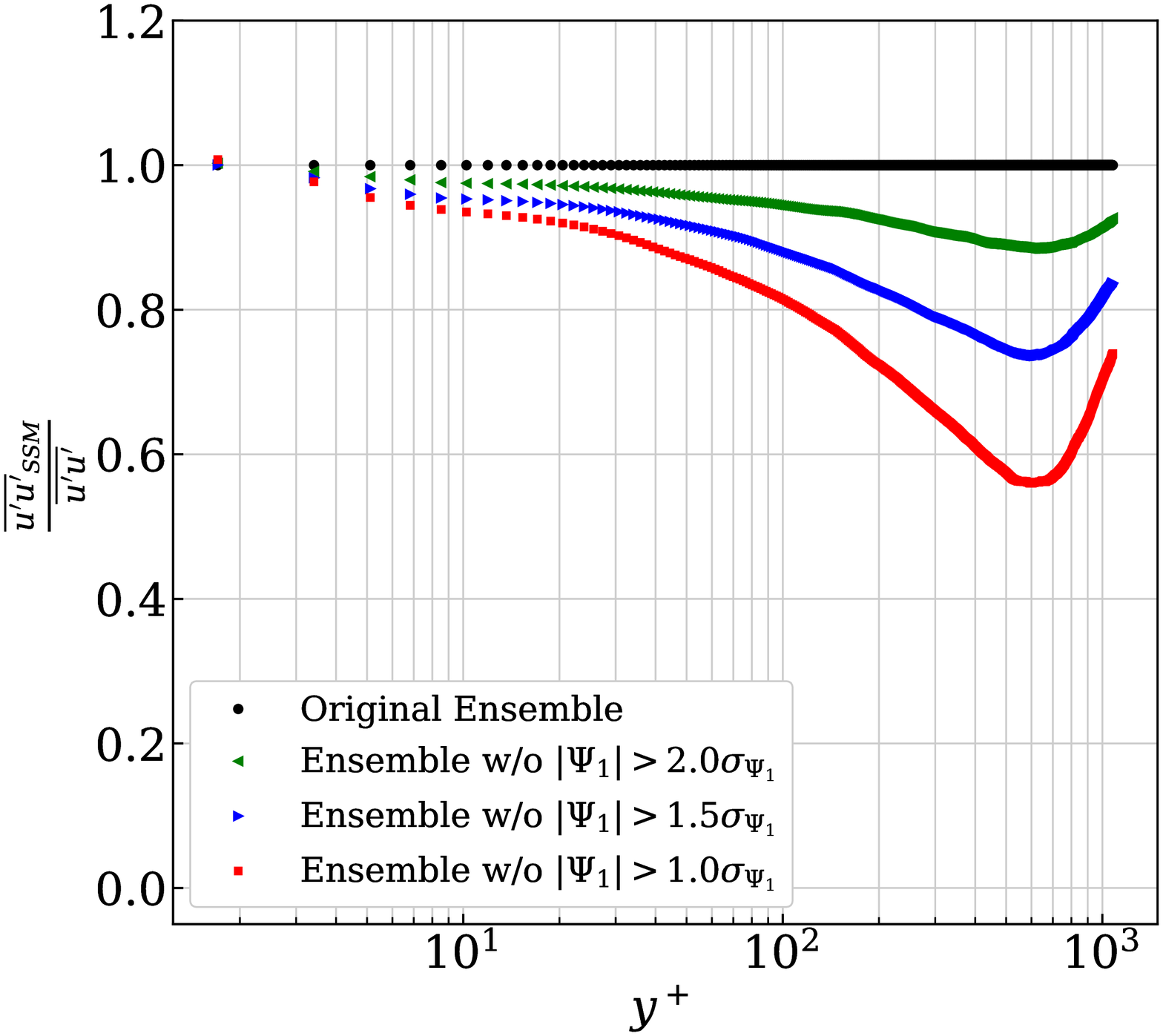} \\
(a) & (b)\\
\includegraphics[trim={1cm 0cm 0cm 0cm},clip,width=0.43\textwidth,height=2.5in]{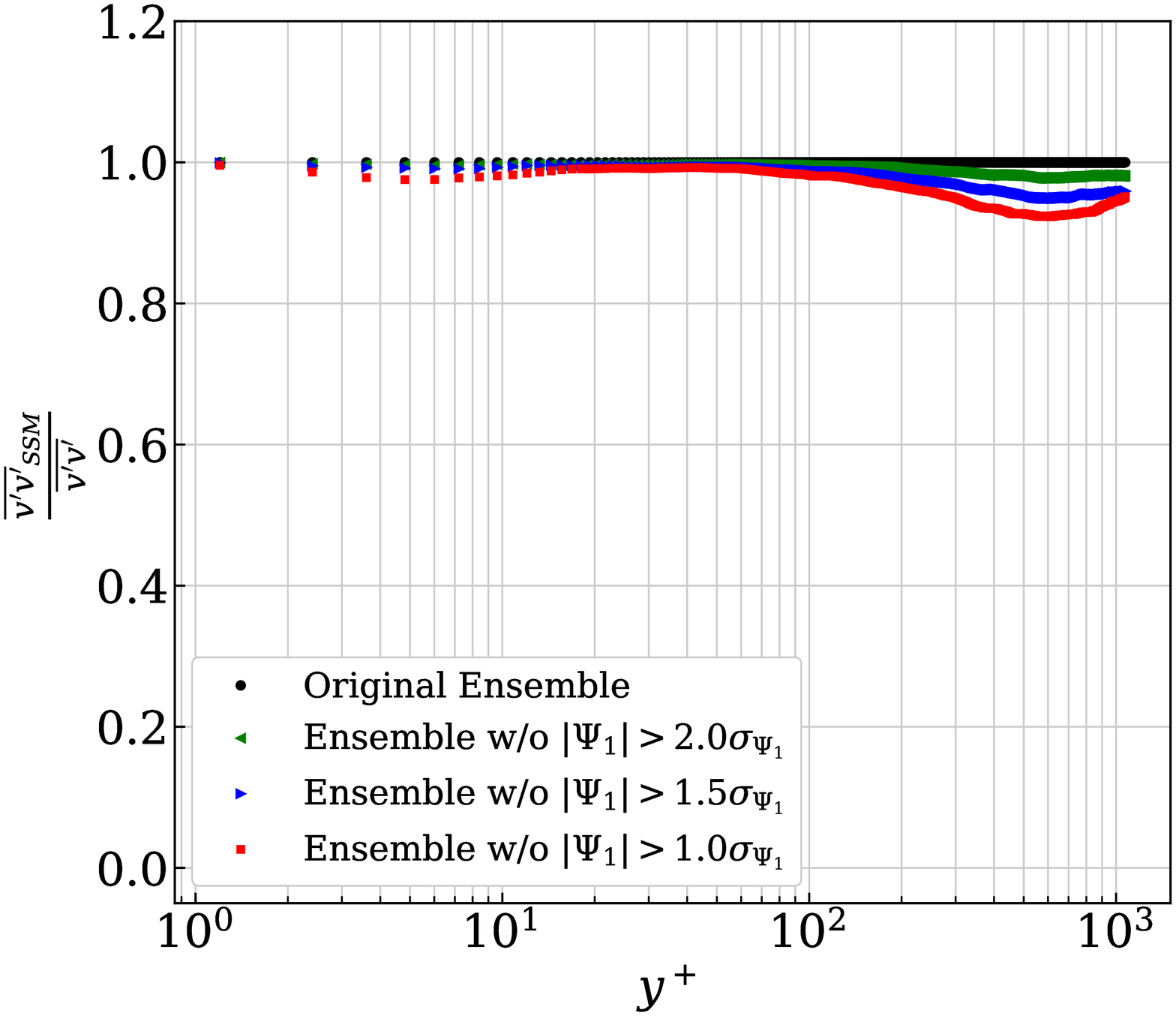} & \includegraphics[trim={2.55cm 0cm 0cm 0cm},clip,width=0.4\textwidth,height=2.5in]{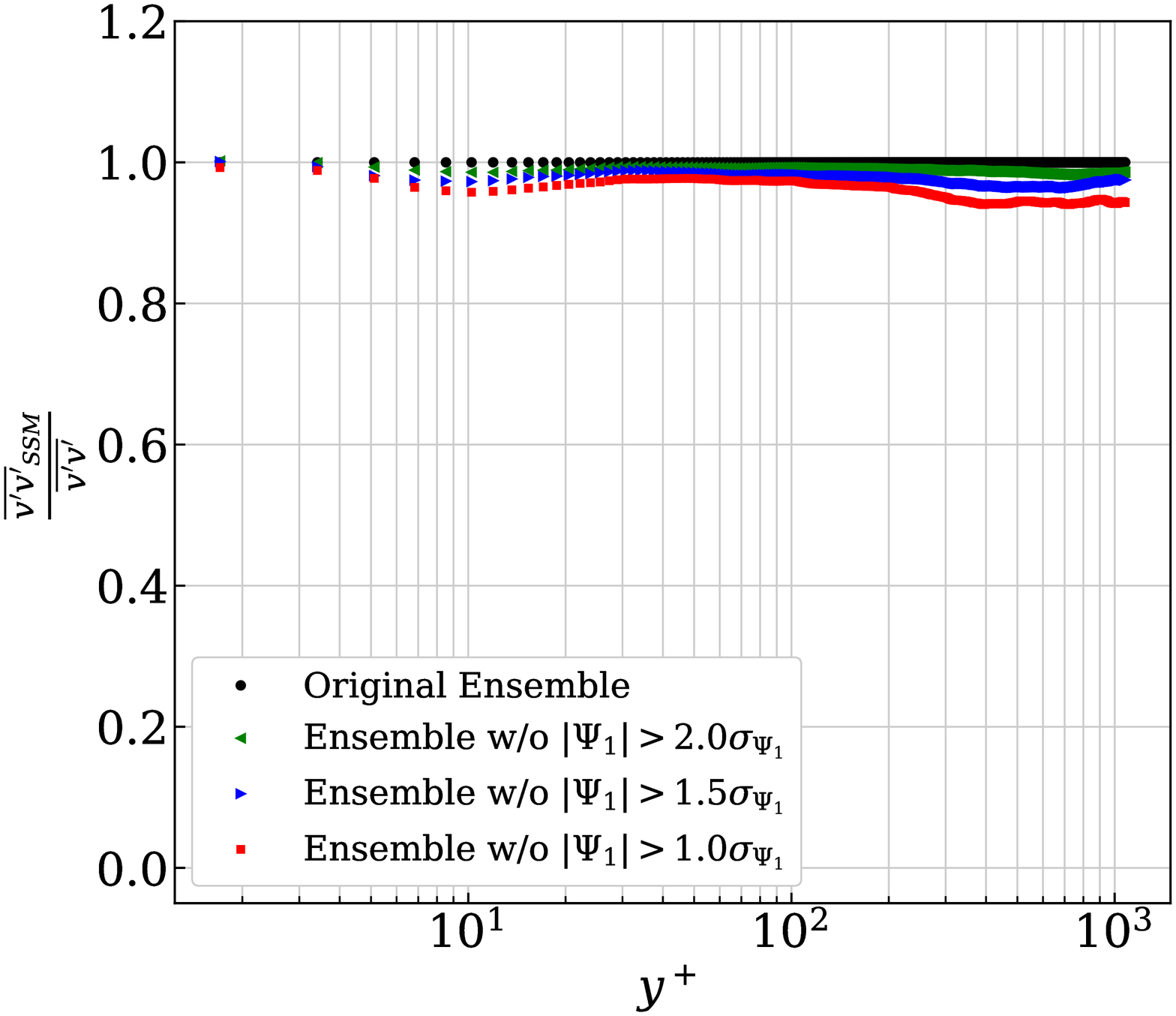} \\
(c) & (d) \\
\includegraphics[trim={1cm 0cm 0cm 0cm},clip,width=0.43\textwidth,height=2.5in]{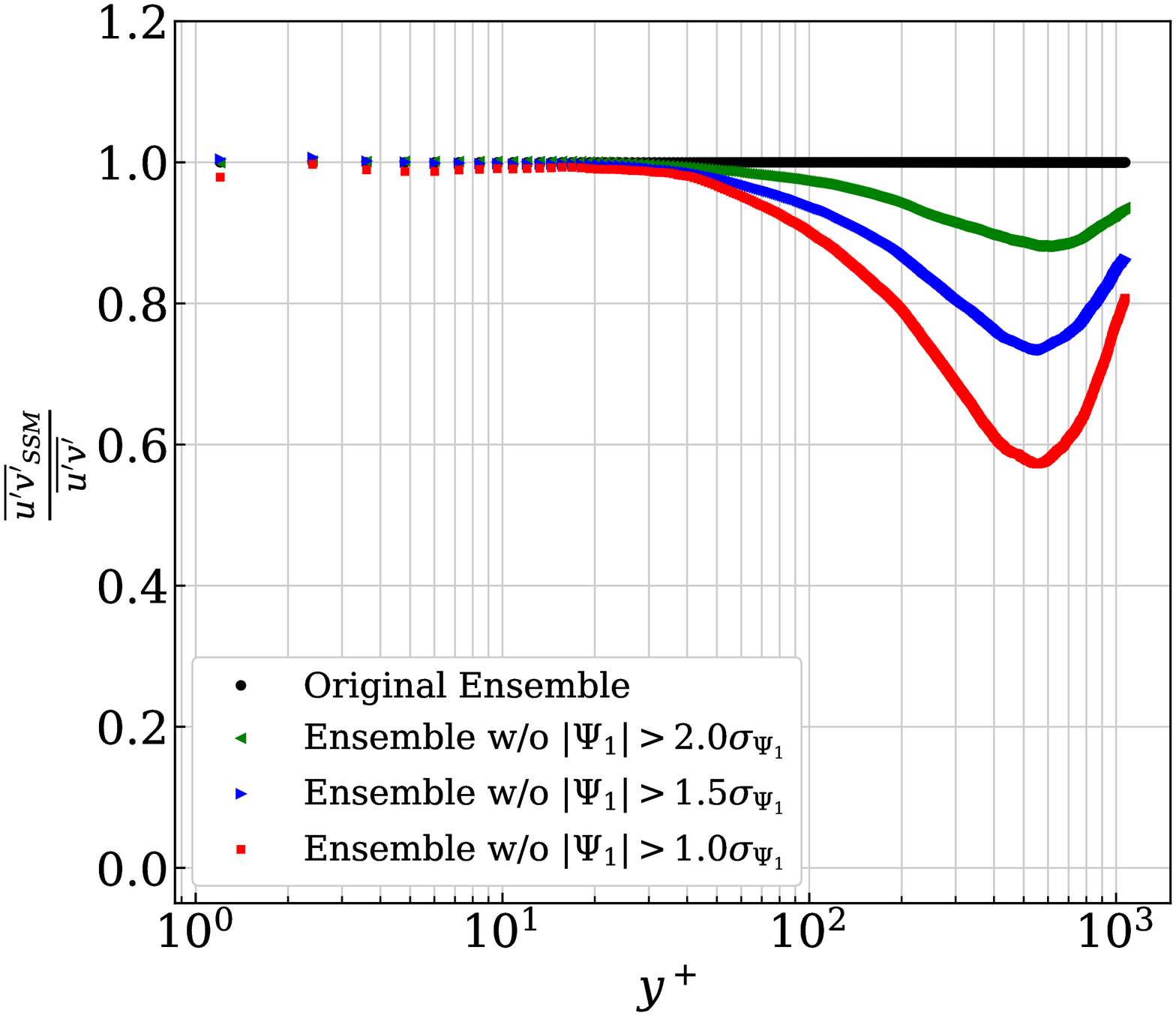} & \includegraphics[trim={2.55cm 0cm 0cm 0cm},clip,width=0.4\textwidth,height=2.5in]{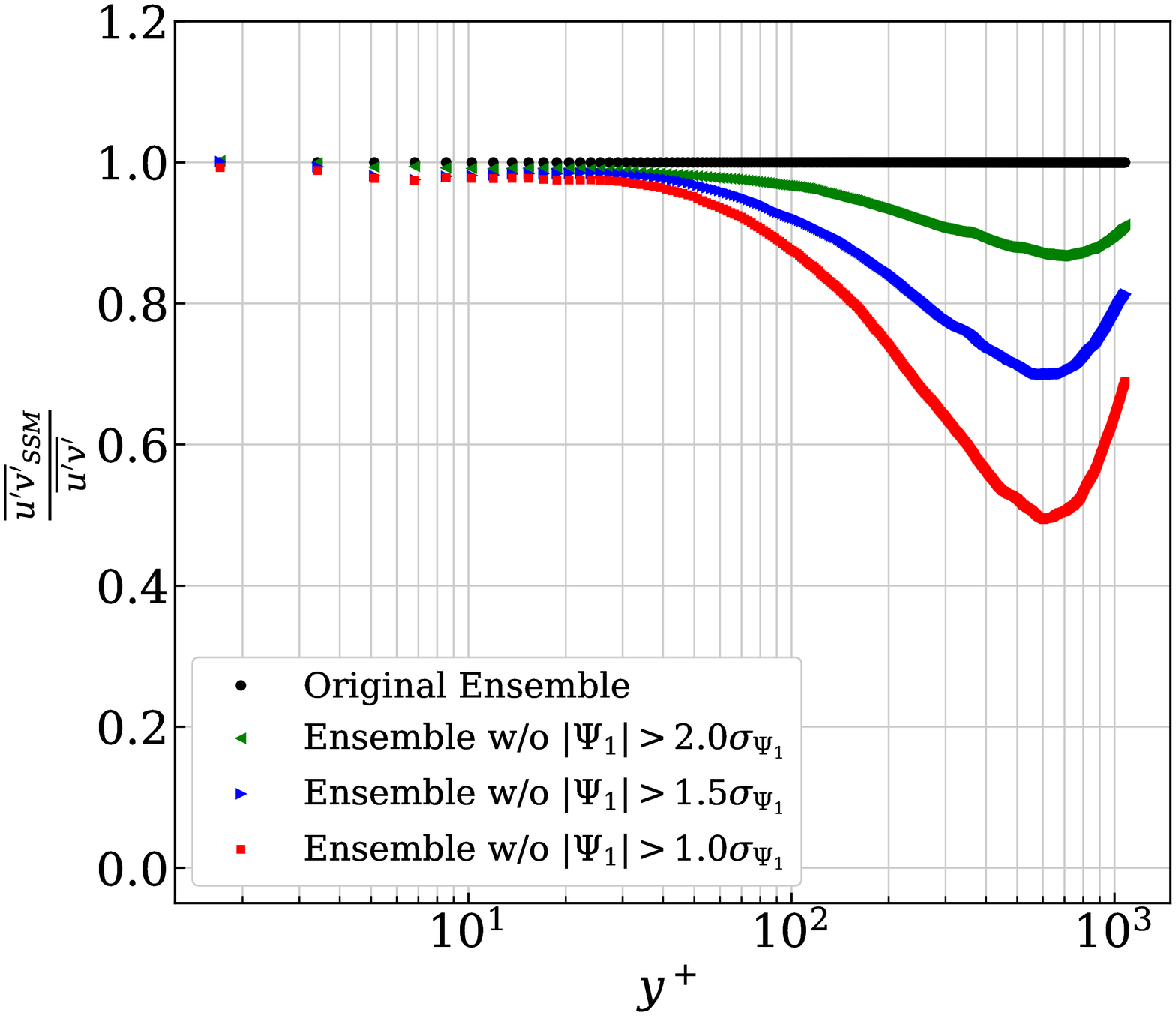} \\
(e) & (f) \\
\end{tabular}
\end{center}
\vspace*{-0.2in}\caption{Reynolds streamwise, wall-normal and shear stresses of in ZPG-TBL (a),(c),(e), and APG-TBL (b),(d),(f), respectively, without the fluctuating velocity fields whose $\psi_1$ values are larger than $K\sigma_{\psi_1}$ where $K=\{1.0,1.5,2.0\}$.
\label{fig:sensitivity_analysis}}
\end{figure}

The sensitivity of Reynolds stresses to the sorted LSMs based on varying the values of $K$ is analysed below. Figure \ref{fig:sensitivity_analysis} shows the profiles of Reynolds stresses in a ZPG- and an APG-TBL, computed from ensembles without the fluctuating velocity fields whose $\psi_1$ values are larger than $K\sigma_{\psi_1}$ where $K=\{1.0,1.5,2.0\}$. These profiles are normalized with the corresponding profiles from the original ensemble. As a larger contribution of the LSMs are removed, the Reynolds stress values are lower than the original ensemble. As shown in figure \ref{fig:sensitivity_analysis}, the effect of a change in $K$ is felt in all of the Reynolds stress profiles in the inner and outer region, but it is more pronounced in the outer region of the Reynolds streamwise and shear stresses. Reynolds wall-normal stresses show very little response to a change in $K$. It is found that the reduction in the Reynolds streamwise stress is the highest, roughly $38\%$ in the ZPG and $43\%$ in the APG, for the smallest value of $K$, which is expected as more instantaneous fluctuating velocity fields with larger $\psi_1$ are excluded from the ensembles even though they are significant contributors to the TKE. This reduction effect becomes less and less as $K$ increases from 1 to 2 and becomes nearly 10\% in the ZPG-TBL and 12\% in the APG-TBL at $K=2$. Although the reduction of the stresses in the inner region after removing LSMs is minimal, it shows that the near-wall scales are affected by the LSMs of the outer region and this effect is also slightly amplified in a mild APG. 

In the Reynolds shear stress, the reduction for different values of $K$ is about 10\% and 15\% larger compared to the Reynolds streamwise stress in the ZPG and APG, respectively. One can conclude that LSMs based on $K=1$ contribute up to 50\% of the Reynolds shear stress in the outer region of the boundary layer.

Profiles of the mean streamwise velocity $U$ and the dominant turbulence production term $-\overline{u'v'}\frac{\partial U}{\partial y}$ are almost insensitive to the change in $K$ and hence, those have not been presented here for the purpose of brevity.

\section{Conclusion}
\label{sec:conclusion}

POD analysis has been performed on the fluctuating velocities of the flow field of a ZPG- and an APG-TBL obtained using high-spatial-resolution PIV. Based on a threshold value $K$, the temporal coefficients of the first POD mode are used to classify the flow field into those with large- and small-scale motions. The fields with large-scale structures are further divided into those with high- and low-momentum events based on the nature of the first POD mode and the values of its temporal coefficients. The conditionally averaged turbulent statistics validate the findings of previous studies in which large-scale motions have significant contributions in the outer region and contribute largely to Reynolds streamwise and shear stresses in a ZPG-TBL. These motions become even more energized in the presence of a mild APG. In the near-wall region, the high- and low-momentum events contribute more and less, respectively, to the turbulent statistics when compared to the original ensemble. Their effect is reversed in the outer region following a cross-over point in the Reynolds stress profiles, which always happens to be above the values averaged over the original ensemble. The two notable differences in the APG-TBL are: (i) high- and low-momentum events are more energized and (ii) the cross-over point moves farther from the wall in the APG-TBL than in the ZPG-TBL.

Removing the velocity fields with the LSMs based on $K=1$, The estimate of the Reynolds streamwise stress and Reynolds shear stress from the remaining velocity fields in the ZPG-TBL are reduced by up to $42\%$ compared to the original ensemble. This reduction effect is enhanced by roughly $10\%$ in a TBL with a mild APG. This further signifies the conclusion that the LSMs are amplified in the presence of a mild APG.

\section*{Acknowledgements}
The authors would like to acknowledge the support of the Australian Government for this research through an Australian Research Council Discovery grant. C. Atkinson was supported by the ARC Discovery Early Career Researcher Award (DECRA) fellowship. The research was also benefited from computational resources provided by the Pawsey Supercomputing Centre and through the NCMAS, supported by the Australian Government. The computational facilities supporting this project included the NCI Facility, the partner share of the NCI facility provided by Monash University through an ARC LIEF grant and the Multi-modal Australian ScienceS Imaging and Visualisation Environment (MASSIVE).

Muhammad Shehzad acknowledges the Punjab Educational Endowment Fund (PEEF), Punjab, Pakistan for funding his PhD research. Bihai Sun and Daniel Jovic gratefully acknowledge the support through an Australian Government Research Training Program (RTP) Scholarship.

\bibliographystyle{elsarticle-harv} 
\bibliography{ref}

\end{document}